\documentclass[aps,pra,onecolumn,tightenlines,floats,superscriptaddress,11pt]{revtex4}
 \usepackage[english]{babel}    
\usepackage[latin1]{inputenc}       
\usepackage[T1]{fontenc}                
\usepackage{graphicx}                       
 \usepackage{lmodern}                        
\usepackage{amssymb,amsmath,float}  

\usepackage[dvipsnames]{xcolor} 

\usepackage{listings}
 \usepackage{fancyhdr}
 \usepackage{url}
 \usepackage{pdfpages} 
 \usepackage{subfigure}
 \usepackage{color}
 \usepackage{tabularx}
 \usepackage{chemfig}
 \usepackage{braket}
 \usepackage{multirow}
\usepackage{bigdelim}
\usepackage{pgfplots}
 \usepackage{bbold}
  \usepackage{pifont}
 \usepackage{tikz}
 \usepackage{pict2e}
\usepackage{calligra}
\usepackage{lineno}

 \newcommand{\fe}{\textbf}
 \newcommand{\om}{\omega}

  \newcommand{\tspace}{\rule{0pt}{2.6ex}}
  \newcommand{\norml}{\textnormal}
  \newcommand{\op}[1]{\mathrm{\hat{#1}}}

\newcommand{\change}[1]{\textcolor{black}{#1}}

\newcommand{\changer}[1]{\textcolor{black}{#1}}
\newcommand{\changem}[1]{\textcolor{black}{#1}}

\newcommand{\changerC}[1]{\textcolor{black}{#1}}
\newcommand{\changerD}[1]{\textcolor{black}{#1}}
\newcommand{\changerE}[1]{\textcolor{black}{#1}}

\begin{document}
\title{Non-adiabatic quantum state preparation and quantum state \change{transport} in chains of Rydberg atoms}

\author{Maike Ostmann}
\affiliation{School of Physics and Astronomy, The University of Nottingham, Nottingham, NG7 2RD, United Kingdom}
\affiliation{Centre for the Theoretical Physics and Mathematics of Quantum Non-equilibrium Systems, The University of Nottingham, Nottingham, NG7 2RD, United Kingdom}
\author{Ji\v{r}\'{i} Min\'{a}\v{r}}
\affiliation{School of Physics and Astronomy, The University of Nottingham, Nottingham, NG7 2RD, United Kingdom}
\affiliation{Centre for the Theoretical Physics and Mathematics of Quantum Non-equilibrium Systems, The University of Nottingham, Nottingham, NG7 2RD, United Kingdom}
\author{Matteo Marcuzzi}
\affiliation{School of Physics and Astronomy, The University of Nottingham, Nottingham, NG7 2RD, United Kingdom}
\affiliation{Centre for the Theoretical Physics and Mathematics of Quantum Non-equilibrium Systems, The University of Nottingham, Nottingham, NG7 2RD, United Kingdom}
\author{Emanuele Levi}
\affiliation{School of Physics and Astronomy, The University of Nottingham, Nottingham, NG7 2RD, United Kingdom}
\affiliation{Centre for the Theoretical Physics and Mathematics of Quantum Non-equilibrium Systems, The University of Nottingham, Nottingham, NG7 2RD, United Kingdom}
\author{Igor Lesanovsky}
\affiliation{School of Physics and Astronomy, The University of Nottingham, Nottingham, NG7 2RD, United Kingdom}
\affiliation{Centre for the Theoretical Physics and Mathematics of Quantum Non-equilibrium Systems, The University of Nottingham, Nottingham, NG7 2RD, United Kingdom}

\begin{abstract}
Motivated by recent progress in the experimental manipulation of cold atoms in optical lattices, we study three different protocols for non-adiabatic quantum state preparation and \change{state transport} in chains of Rydberg atoms. The protocols we discuss are based on the blockade mechanism between atoms which, when excited to a Rydberg state, interact through a van der Waals potential, and rely on single-site addressing. Specifically, we discuss protocols for efficient creation of an antiferromagnetic GHZ state, a class of matrix product states including a so-called Rydberg crystal and for the \change{state transport} of a single-qubit quantum state between two ends of a chain of atoms. We identify system parameters allowing for the operation of the protocols on timescales shorter than the lifetime of the Rydberg states while yielding high fidelity output states. We discuss the effect of positional disorder on the resulting states and comment on limitations due to other sources of noise such as radiative decay of the Rydberg states. The proposed protocols provide a testbed for benchmarking the performance of quantum information processing platforms based on Rydberg atoms.

\end{abstract}

\maketitle

\section{Introduction}

Cold atoms held in optical traps constitute an invaluable tool in the quest for quantum information processing (QIP) and simulation of many-body physics in the quantum regime \cite{Bloch_2008,Bloch_2012}. Rydberg gases, i.e. atoms excited to high principal quantum number states, are of particular interest as the strong interaction between atoms in the Rydberg states can be exploited for various quantum information processing tasks \cite{a_Saffman_RMP_10,Weimer_2010,Saffman_2016}. Experimental progress in manipulating these Rydberg atoms now allows simulating quantum Ising Hamiltonians \cite{Labuhn_2015}, the adiabatic preparation of the ground states thereof \cite{Schauss_2015}, efficient entanglement creation \cite{Jau_2016} or the implementation of quantum gates \cite{Isenhower_2010,Maller_2015}. Some of these advances rely on the use of optical tweezer arrays, which permit the creation of various lattice geometries \cite{Nogrette_2014} and were recently used to deterministically obtain an optical lattice with close-to-unit filling \cite{Barredo_2016,Endres_2016}. Importantly, techniques allowing for addressing a single atom in such arrays have been developed \cite{Labuhn_2014,Bloch_2016,Greiner_2009,Zwierlein_2015} opening new possibilities for \emph{non-adiabatic} quantum state engineering, which might help to overcome limitations imposed by the required timescales for adiabatic procedures, where detrimental relaxation effects may become important \cite{a_Petrosyan_arxiv_16}. First steps in this direction were taken e.g. in Ref. \cite{Cui_2017}
which considered optimal control techniques for creation of ferromagnetic GHZ, crystalline or Fock superposition quantum states in Rydberg atoms.\\

 Building on the capabilities of optical tweezer arrays with Rydberg atoms and single site addressing for QIP, we consider GHZ and matrix product state (MPS) engineering and quantum \change{state transport} in a one-dimensional geometry.
 This particular choice is motivated by the fact that all three examples play a fundamental role in QIP and constitute an ideal benchmark in order to assess the performance of the experimental platform we consider - \changerD{an array of Rydberg atoms} - for our theoretical and numerical analysis. Specifically, the GHZ state serves as a reference in quantum estimation theory yielding the Heisenberg scaling \cite{a_Giovannetti_Science_04}. Various proposals exist in the literature for the creation of GHZ states \cite{a_Molmer_PRL_99,a_Saffman_RMP_10,a_Wang_arxiv_16} some of which have been realized experimentally, for example using ultracold ions \cite{a_Monz_PRL_11}. Similarly, MPSs play a central role in classical simulations of quantum Hamiltonians in one dimension \cite{Schollwoeck_2005, Perez_2007, Schollwoeck_2011} and are naturally realized as ground states of some finite-range interaction spin chains \cite{a_Lesanovsky_PRL_11, a_Lesanovsky_PRL_12, Levi_2014} which are related to the problem of classical hardcore dimers \cite{a_Rokhsar_PRL_88}. For that reason we refer to the class of MPS considered in this article as dimer-MPS. \changerE{Importantly, the dimer-MPS feature the so-called Rydberg crystal as a special case \cite{Schachenmayer_2010,VanBijnen_2011,Pohl_2010,Schauss_2015}.} 
 \change{Finally, faithful transport of a quantum state between different nodes of a quantum network is an essential requirement for QIP schemes such as quantum computation \cite{a:Bennett_00}. Various methods to achieve quantum state transport between spatially separated qubits have been proposed. These include schemes based on atoms connected through an optical link \cite{a:Cirac_97} or Rydberg atoms, where the transport is achieved through interaction between the Rydberg atoms and atomic ensembles which communicate through a photon exchange \cite{a:Saffman_05}.
}

In this paper, the QIP is based on the so-called ``Rydberg blockade'' mechanism which relies on the strong repulsive interaction
between atoms excited to a Rydberg state \cite{a_Saffman_RMP_10}. We first introduce the protocols for GHZ state
and {dimer-MPS} generation and quantum \change{state transport}
in \change{the idealized limit of perfect blockade} in Sec. \ref{sec:ideal_case}. 
\change{In this regime the blockade mechanism can be effectively described by a three-body Hamiltonian which constitutes the basic building block 
of the studied protocols.}
Next, we investigate the influence of more realistic conditions, such as the non-perfect blockade due to the finite value
of the interaction energy and the tails of the interaction or the positional disorder of the atoms held in optical
tweezers \cite{a_Marcuzzi_PRL_17} in Sec. \ref{sec:Imperfections}. 
\change{There, we relax the requirement of strict blockade and consider instead an evolution guided by a
more realistic system Hamiltonian including a van der Waals interatomic potential. This allows us to verify the predictions of the effective 
description of Sec. \ref{sec:ideal_case}.}
To this end we evaluate the fidelity of the produced states with respect to the
target as a function of various parameters, such as the Rabi frequency of the laser pulses, interaction strength,
length and parity of the chain or the strength of the disorder. We summarize and discuss the results in Sec. \ref{sec:conclusion}.


\section{Setup, state preparation and $\change{\text{STATE TRANSPORT}}$ protocols}
\label{sec:ideal_case}

\begin{figure}
\label{setup_N_2}
\begin{minipage}{0.4\textwidth}
   \includegraphics[height=5cm]{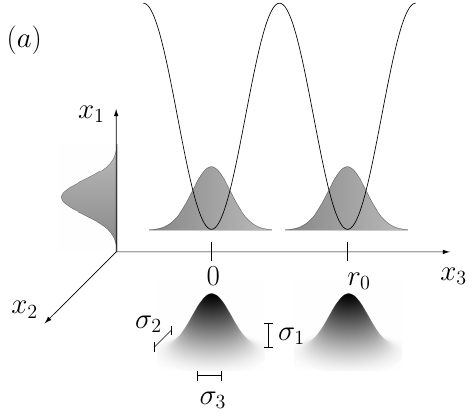} 
\end{minipage}
\begin{minipage}{0.2\textwidth}
 \centering
 \vspace{4em}
 \includegraphics[height=2cm]{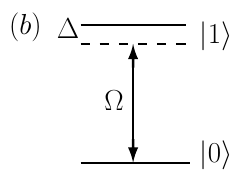}
 \vspace{3em}
 \label{2_level_sys}
  \end{minipage}
\begin{minipage}{0.2\textwidth}
 \centering
  \vspace{4em}
 \includegraphics[height=2.3cm]{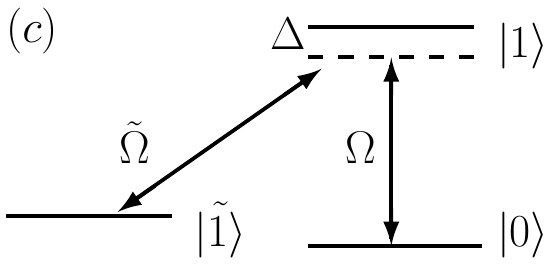}
  \vspace{1em}
 \label{3_level_sys}
 \end{minipage} 
 
  \caption{(a): Setup for $N~=~2$ atoms. The optical traps are arranged along the $x_3$ direction and are separated by $r_0$. The position of the atoms is spread with uncertainty $\sigma_i$ around their equilibrium values \cite{a_Marcuzzi_PRL_17}. 
(b): Level scheme for an effective two level system, where the state $\ket{0}$ is coupled to the Rydberg state $\ket{1}$ by a laser with Rabi frequency $\Omega$. (c): Energy levels of an effective three-level system as it is assumed in the GHZ protocol, where $\tilde \Omega$ couples the Rydberg state $\ket{1}$ to another hyperfine state $\ket{\tilde1}$.
}
\label{fig:scheme}
\end{figure}

We consider a one-dimensional chain (open boundaries) along the $x_3$-direction of $N$ optical traps, each occupied with a single atom. A diagram of the setup for the case of two atoms is shown in Fig. \ref{fig:scheme}a. The optical traps are separated by equal spacings $r_0$ so that the position of the $k$-th atom reads $\fe{r}_k~=~(0,0,k r_0)$. Each atom is described as an effective two level system, where the electronic ground state $\ket{0}$ is coupled to the highly excited Rydberg state $\ket{1}$ via a laser pulse with Rabi frequency $\Omega$ as depicted in Fig. \ref{fig:scheme}b. For later convenience, we account for the presence of a second hyperfine ground state $\ket{\tilde1}$ coupled to the Rydberg state via a second laser with different polarization and Rabi frequency $\tilde{\Omega}$, see Fig. \ref{fig:scheme}c.

Considering that the atoms in Rydberg states interact through the van der Waals potential, the Hamiltonian of the system is given by (in the rotating wave approximation)
\begin{align}
 \op{H} = \sum_{k = 1}^N \op{h}_k(\Omega_k,\Delta_k) + \op{H}_{\rm int},
 \label{eq:full_hamil}
\end{align}
where
\begin{subequations}
\begin{align}
 \op{h}_k(\Omega_k,\Delta_k) &= \Omega_k \, \op{\sigma}_y^{(k)} \, + \, \Delta_k \, \op{n}_k \\
 \op{H}_{\rm int} &= \frac{V_0}{2} \sum_{\substack{k,m = 1\\ m \ne k}}^N \, \frac{\op{n}_m\, \op{n}_k}{|k -m|^6}. \label{eq:Hint}
\end{align}
	\label{eq:full_hamil_piecewise}
\end{subequations}
Here, $\op \sigma_y^{(k)} = \norml{i}\,(\ket{1_k}\bra{0_k}- \ket{0_k}\bra{1_k})$, $\op{n}_k = \ket{1_k}\bra{1_k}$, the parameter $\Delta_k$ is the relative detuning of the laser from the resonant transition between 
the states $\ket{0_k}$ and $\ket{1_k}$ and $V_0 = C_6/r_0^6$ is the interaction strength proportional to the van der Waals coefficient $C_6$. Note that in (\ref{eq:full_hamil}) we have allowed for site-dependent Rabi frequencies and detunings $\Omega_k,\Delta_k$.

In the regime where $V_0 \gg \Omega_k,\; \forall k$, the driving $ \propto\Omega_k$, which induces spin flips in the $\op \sigma_z^{(k)}$ basis, cannot overcome the energy cost of having two neighboring Rydberg excitations. This is commonly referred to as the ``Rydberg blockade'' mechanism. In this limit, it becomes convenient to adiabatically eliminate blockaded processes: applying a unitary transformation
$\op{U}={\rm exp}[-\norml{i}\,t\,V_0 \sum_{k=1}^{N-1} \op{n}_k \op{n}_{k+1}]$ on (\ref{eq:full_hamil}), neglecting terms oscillating at the frequencies $V_0$ and interactions beyond nearest-neighbor and considering resonant excitation $\Delta_k=0,\; \forall k$, one can derive an effective three-body Hamiltonian making the blockade mechanism manifest \cite{a_Lesanovsky_PRL_11}:
\begin{equation}
 \op H = \sum_{k=1}^N ~ \op{h}_k^{3\,\norml{body}}.
  \label{eq:hamil_lopt}
\end{equation} 
Here, $\op{h}_k^{3\,\norml{body}} = \Omega_k~ \op P_{k-1} \op{\sigma}_y^{(k)} ~ \op P_{k+1}$ and $\op{P}_k = (1- \op{n}_k)$
are the projectors on ground-state atoms. The associated unitary evolution corresponding
to the application of a laser pulse of duration $t_k$ and area $A_k=\Omega_k t_k$ on the $k$-th atom on the $\ket{l}-\ket{l'}$ transition reads
\begin{align}
  \op{U}^{ll'}_{{k}}(A_k) = \exp(-\norml{i}\, t_k\, \op{h}_k^{3\,\norml{body}} ) = \mathbb{1} -\op{P}_{k-1}\op{P}_{k+1} + \op{P}_{k-1}\op{P}_{k+1}\norml{e}^{-\norml{i}\, t_k\, \Omega_k\, \op{\sigma}_y^{(k)}}
  \label{eq:U 3body}
\end{align}
and represents the basic building block of the protocols studied in this paper \footnote{In principle more general unitaries can be obtained by considering $\hat{h}_k({\mathbf \Omega}_k,\Delta_k) = \Omega^x_k \, \op{\sigma}_x^{(k)}  + \Omega^y_k \, \op{\sigma}_y^{(k)}\, + \, \Delta_k \, \op{n}_k$.}. \changerC{The indices $ll' \in \{01,1\tilde{1} \}$ on the left-hand side of (\ref{eq:U 3body}) label the basis in which the operators are expressed. For example, ${\rm \hat{U}}^{01}_k$ means that the operators ${\rm \hat{P}}_k = 1- \hat{n}_k$ and $\hat{\sigma}_y^{(k)}$ on the right-hand side of (\ref{eq:U 3body}) act upon the $\{\ket{0},\ket{1}\}$ basis.} We note that for $A_k~=~\pi$, the unitary (\ref{eq:U 3body}) corresponds to the Toffoli gate (which we recall in Appendix \ref{app:Toffoli}) with $k$ the target and $k-1,\,k+1$ the control atoms. In what follows we shall refer to the preparation procedure as \emph{non-adiabatic} meaning that the state of the system evolves in a step-wise manner after every application of a gate of the form (\ref{eq:U 3body}) [or any other local gate] and in general is not an eigenstate of the Hamiltonian of the system, {Eq. (\ref{eq:full_hamil}). This has to be contrasted with adiabatic protocols, where the final state is the ground state of a Hamiltonian whose parameters are adiabatically deformed, starting from an initial ground state that is easy to prepare.}

Before we proceed with the introduction of the state preparation and \change{state transport} protocols, a comment is in place. The reason for considering the auxiliary hyperfine state $\ket{\tilde{1}}$ is to avoid undesirable effects such as spontaneous emission from atoms in the Rydberg levels or {atomic loss and dephasing due to mechanical forces acting on the atoms (van der Waals repulsion \cite{Thaicharoen_2015,Faoro_2016} and collisions with the background gas). This can be achieved in the three-level scheme when the atomic population is transferred from the Rydberg level $\ket{1}$, after it has been used to implement a particular gate, as fast as possible to the stable hyperfine state $\ket{\tilde{1}}$.} However, in order to reduce the complexity of the experiments {(which may be important for first proof-of-principle demonstrations)} and to further reduce the number of applied gates, we will consider the two-level configuration as well, where only the ground and Rydberg levels $\ket{0},\ket{1}$ are involved. More specifically, we describe the use of both the three and two-level schemes on the example of the GHZ state preparation protocol in Sec. \ref{sec:GHZ protocol def} and discuss the differences between the two schemes in Sec. \ref{GHZ-vdW}. For the reasons mentioned above, we limit the discussion of the dimer-MPS preparation and \change{state transport} protocols to the two-level scheme only.


\subsection{\changerC{GHZ state preparation}}
\label{sec:GHZ protocol def}
We consider the non-adiabatic preparation of antiferromagnetic GHZ states of the form
\begin{align}
 \ket{\norml{GHZ}} = \frac{1}{\sqrt{2}}\, 
 \left(\, \ket{0_1\, \tilde{1}_2\,0_3 \ldots \tilde{1}_N} + \ket{\tilde{1}_1\, 0_2\,\tilde{1}_3 \ldots 0_N }\,\right)\, .
 \label{eq:GHZ}
\end{align}
They are robust with respect to global noise, such as magnetic or electric field fluctuations on length-scales larger than the length of the chain, as the two components of the state are energetically degenerate: using encoding in the basis $\ket{0},\ket{\tilde{1}}$, this is only strictly true when $N$ is even as one has the same number of excitations (atoms in the state $\ket{\tilde{1}}$) in both components of the GHZ state [the two terms on the rhs of (\ref{eq:GHZ})].

Initially all atoms are prepared in the ground state $\ket{\Psi_{\norml{in}}}~=~\ket{0\,0\,0\ldots0}$. 
First we apply a $\pi/2$ pulse $\op{U}^{01}_{1}\left({\frac{\pi}{2}}\right)$ on the first atom to generate a superposition state
\begin{align}
 \ket{\Psi_1} = \op{U}^{01}_{1}\left(\frac{\pi}{2}\right)\, \ket{\Psi_{\norml{in}}}
 = \frac{1}{\sqrt{2}}(\ket{0 \, 0\, 0\, \ldots\, 0} +\ket{1 \, 0\, 0\,\ldots\,0})\,.
 \label{eq:GHZ prep 1}
\end{align}
This is followed by the application of a $\pi$-pulse $\op{U}^{01}_{{{2}}}(\pi)$  on the second atom
\begin{align}
  \ket{\Psi_2} = \op{U}^{01}_{{2}}(\pi)\, \ket{\Psi_{\norml{1}}}
 = \frac{1}{\sqrt{2}}(\ket{0\,1\,0\,\ldots\,0} +\ket{1\,0\,0\,\ldots\,0})\,.
\end{align}
Note that, due to the blockade mechanism, the second term on the rhs of (\ref{eq:GHZ prep 1}) is not affected by the second pulse. We next go back to the first atom and apply a $\pi$-pulse on the $\ket{\tilde{1}}-\ket{1}$ transition in order to transfer any population in its Rydberg level to the hyperfine state $\ket{\tilde{1}}$ so that
\begin{align}
  \ket{\Psi_3} = \op{U}^{1\tilde{1}}_{{1}}(\pi)\, \ket{\Psi_{\norml{2}}}
 = \frac{1}{\sqrt{2}}(  \ket{0\,1\,0\,\ldots\,0} -\ket{\tilde{1}\,0\,0\ldots\,0})\,.
\end{align}
This is followed by the application of $\op{U}^{01}_{3}(\pi)$ on the third atom and $\op{U}^{1\tilde{1}}_{{2}}(\pi)$ on the second atom and so forth until the end of the chain is reached after $2N-1$ gates (unitaries) have been applied. The procedure is summarized in Table \ref{tab:GHZ_protocol}. We note that a similar proposal for the preparation of a ferromagnetic GHZ state was put forward recently in \cite{a_Wang_arxiv_16}.

With the GHZ state preparation protocol just described, the Rydberg state $\ket{1}$ appears neither in the initial nor final state, Eq. (\ref{eq:GHZ}). It is simply exploited to implement the constrained spin flipping which allows to reconstruct the GHZ pattern in the two components of the state. \changerE{A simplified two-level version of this protocol, where only levels $\ket{0}$ and $\ket{1}$ are used, is obtained simply by neglecting the transfer from the Rydberg to the hyperfine state, i.e. omitting the step (iv) in Table \ref{tab:GHZ_protocol}.}

\begin{table}
\centering
     \begin{tabular}{ |c| l |l|l | }
     \hline
	       & \multicolumn{1}{c|}{GHZ}                      & \multicolumn{1}{c|}{dimer-MPS}                                &  \multicolumn{1}{c|}{\change{Transport}}    \\
	  \hline
$\ket{\Psi_\norml{in}}$     & $\ket{0\,0\,0\ldots0}$           &  $\ket{0\,0\,0\ldots0}$                                 &  $\ket{\psi\,0\,0\ldots0}$\rule{0pt}{3.5ex} \\
	  (i)  & apply $\op{U}^{01}_{{1}}(\frac{\pi}{2})$      &   set counter to $k = 1$                                & set counter to $k = 1$ \rule{0pt}{3.5ex}    \\
               &                                               &                                                         &          \\ 
               &                                               &                                                         &\\
         (ii)  & set counter to $k = 1$                        &  apply pulse of area $A_k$ on                                   & apply $\pi$-pulse \rule{0pt}{3.5ex}   \\ 
               &                                               & atom $k$ [see Eqs. (\ref{eq:MPS BC})]   & on atom $k+1$  \\
         (iii) & apply $\op{U}^{01}_{{{k+1}}}(\pi)$            & $k  \Rightarrow k+1$                                    & apply $\pi$-pulse\rule{0pt}{3.5ex}  \\
               &    if $k$ = $N$ do nothing                    &  go back to \change{(ii)}                                       & on atom $k$ \\
               &                                               & till $k$ = $N$ then stop                                &\\
          (iv) &  apply $\op{U}^{1\tilde{1}}_{{{k}}}(\pi)$     &                                                         & $k  \Rightarrow k+1$\rule{0pt}{3.5ex} \\
               &                                               &                                                         & go back to (ii) \\
               &                                               &                                                         & till $k$ = $N$ then stop  \\
           (v) & move to next atom                             &                                                        & \changerE{apply $\norml{i}^{N-1} \hat{\sigma}_y^{(N)}$ for $N$ even,} \rule{0pt}{3.5ex}  \\
               &  $k  \Rightarrow k+1$                         &                                                        & \changer{see end of Sec. \ref{sec:Teleportation ideal}} \\
               & repeat (iii) - (v)                            &                                                        &   \\
               & till $k$ = $N$                                &                                                        &  \\
               \hline
 $\#$ of applied pulses      &  $2N-1$    	                               & $N$                                                    & $2N-2$\rule{0pt}{3.5ex}\\
 \hline
    \end{tabular}
    \caption{Pulse sequence for the preparation of a GHZ state (left column), the dimer-MPS state (middle column) and the \change{state transport} (right column) in a chain of $N$ Rydberg atoms.
    In the case of the GHZ state, the Rydberg atoms are described as an effective three-level system. However, the GHZ protocol can also be adapted to a two-level description of the atoms by neglecting step (iv). 
    For the dimer-MPS as well as the \change{state transport} protocol, each atom is approximated as a two-level system. $\ket{\Psi_\norml{in}}$ is the initial state for the particular protocol.}
    \label{tab:GHZ_protocol}
\end{table}


\subsection{Dimer-MPS preparation}
\label{sec:MPS}

\begin{figure}
\begin{subfigure}

\includegraphics[height=3cm]{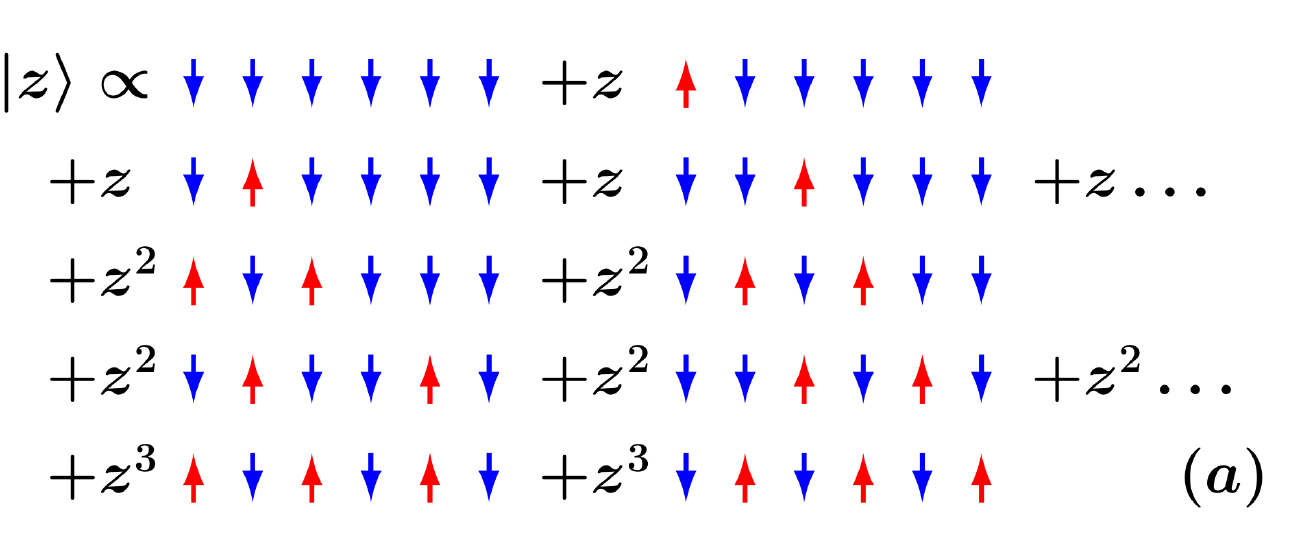}
 \label{RK_GS_pic}
\end{subfigure}
\begin{subfigure}

    \includegraphics[height=3.5cm]{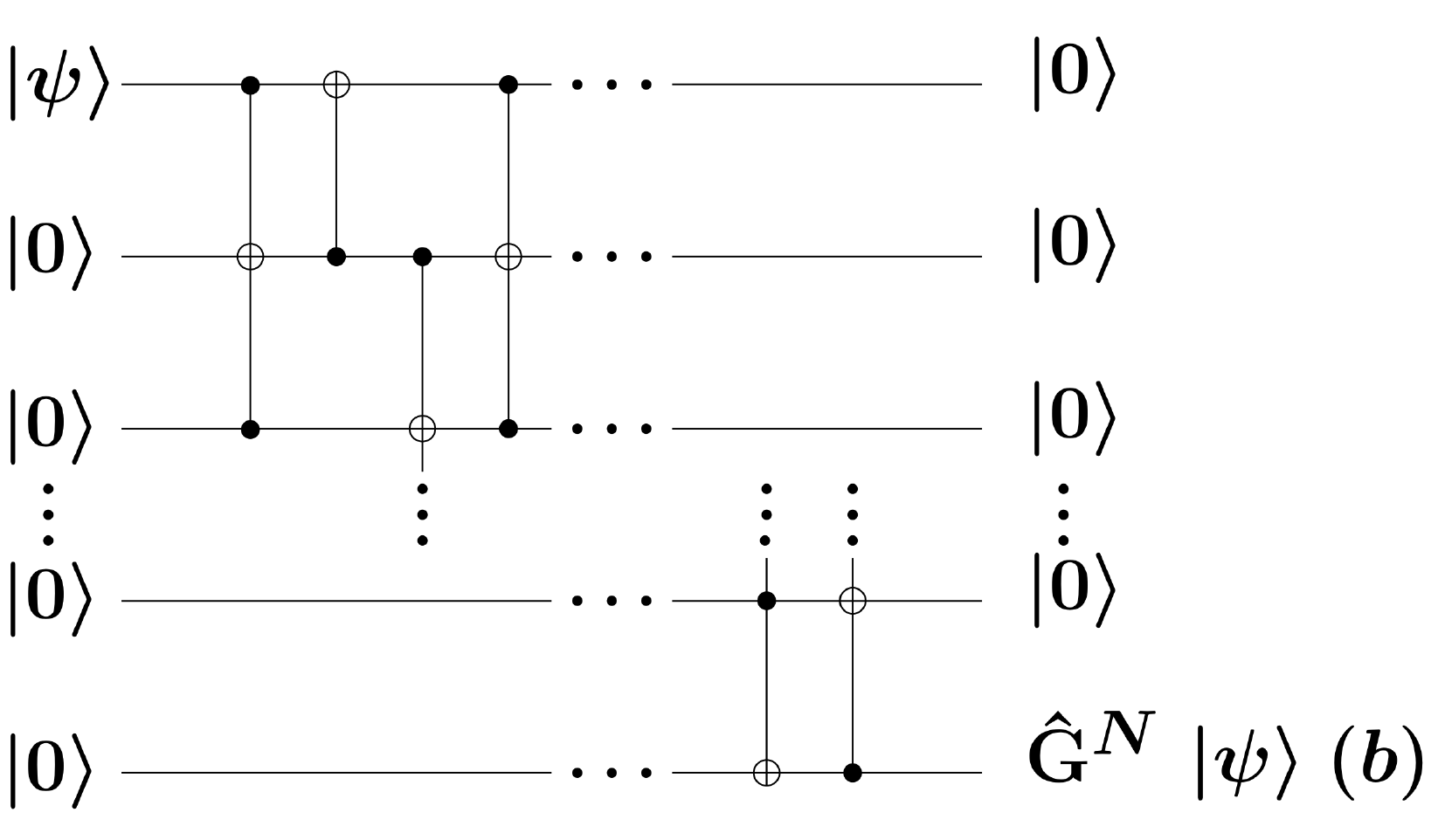}
\label{coherent_teleport_Nqubit}
\end{subfigure}
\caption{(a): Representation of the dimer-MPS for a chain of $N$~=~6 atoms. This state is a superposition of all possible configurations without neighboring Rydberg excitations. The parameter $z$ is weighting the number of excitations in the particular component of the superposition state. (b): Circuit for the \change{transport} of a single qubit state $\ket{\psi}$ from the first to the $N$th qubit in the chain. All qubits expect the first one, are prepared in the ground state. The CNOT gate as well as the Toffoli gate correspond to a $\pi$-pulse with Rabi frequency $\Omega$ applied to the particular atom. In the post processing, the $\op G^{N}$-gate is applied to the state on the $N$th qubit, in order to obtain the correct state $\ket{\psi}$.}

\label{fig:MPS_n_teleport}
\end{figure}

It seems natural to explore the effective three-body interaction $\hat{h}_k^{3\rm{body}}$ used in the GHZ state preparation for creation of various other quantum states. Specifically, we note that the perfect blockade and the associated Hamiltonian (\ref{eq:hamil_lopt}) allow in principle for the creation of all possible configurations \change{which are compatible with the dimer-MPS protocol.
More precisely, in the blockade regime, simultaneous excitations of adjacent atoms are strongly suppressed, which 
confines the dynamics to subspaces where the number of neighboring excitations remains constant. In particular, 
the subspace we work in displays all Rydberg excitations separated by at least on side.} 
In this section we describe how to produce a specific example of such quantum state which is defined as
\begin{align}
 \ket{z} = 
 \frac{1}{\sqrt{Z_z}}\, \prod_{k=1}^N ( \mathbb{1}\, + \,z\,\op{P}_{k-1}\, \op \sigma_k^+ \,\op P_{k+1})\, \ket{0\, \ldots \, 0}\,.
 \label{eq:ideal_z}
\end{align}
Here, $z$ is a real number parametrizing the state, $\op P_k = (1- \op{n}_k) = \ket{0_k}\bra{0_k}$ is the projector on the atomic ground state introduced after Eq.~\eqref{eq:hamil_lopt} and $Z_z$ is an overall normalization constant ensuring $\braket{z|z}=1$. The state (\ref{eq:ideal_z}) is a superposition of all possible configurations without adjacent Rydberg excitations, where each configuration is weighted by $z^n$ with $n$ the total number of (isolated) excitations in it. An illustration of $\ket{z}$ is provided in Fig.~\ref{fig:MPS_n_teleport}a.
%
Specifically, for $z = 0$, $\ket{z}$ is the product state $ \ket{0\,0\,0\,\ldots\,0}$, without any entanglement.
For $z$ = 1, all possible components are equally weighted, whereas \change{for $z \gg 1$} the ones with higher number of excitations dominate.
Note that for odd $N$ and \change{$z \rightarrow \infty$, the corresponding dimer-MPS} state is a single antiferromagnetic configuration where all odd sites,
including the first and the last, have an excited atom. \change{Such an alternating configuration constitutes an example of a so-called Rydberg crystal.
However, this state is completely separable and holds no entanglement at all. For the scope of this work, it is
therefore convenient to focus on the complementary case:} for even $N$, the major contribution will instead be given by the $N/2 + 1$ configurations with $N/2$ excitations.

It has been shown in \cite{a_Lesanovsky_PRL_12} that the state (\ref{eq:ideal_z}) admits a matrix product representation
\begin{align}
 \ket{z} = \frac{1}{N}\sum_{i_1,\ldots,i_N = 0,1}[\,\vec{l}\, \hat{X}_{i_1}\, \hat{X}_{i_2}\,\ldots  \hat{X}_{i_N}\, \vec{r}\,] \ket{i_1\, i_2\, \ldots i_N}\, ,
 \label{eq:MPS z}
 \end{align}
 with $\op{X}_0 = (\mathbb{1}-\op{n}) +z\,\op{\sigma}_-$, $\op{X}_1 = \op \sigma_+$ two $2\times 2$ matrices, and ladder operators
 $\op \sigma_{\pm} = (\op\sigma_x \pm \norml{i}\op \sigma_y)/2$. The vectors $\vec{l} = (z,1)$ and $\vec{r} = (0,1)$
 are  included to \changer{impose} the correct boundary conditions. 
 
Furthermore, the same construction leading to (\ref{eq:ideal_z}) can be generalized to the case of a blockade extending over $R$ sites (i.e. an excitation prevents its first $R$ neighbors from being excited) \cite{Levi_2014}. The analog of the state (\ref{eq:ideal_z}) then reads
\begin{align}
 \ket{z} = 
 \frac{1}{\sqrt{Z_z}}\, \prod_{k=1}^N ( \mathbb{1}\, + \,z\,\op{P}_{k,\norml{left}}\, \op \sigma_k^+ \,\op P_{k,\norml{right}})\, \ket{0\, \ldots \, 0}\,,
 \label{eq:ideal_zR}
\end{align}
where $\op P_{k,\norml{left}} = \prod_{j=1}^R \op P_{k-j}$, $\op P_{k,\norml{right}} = \prod_{j=1}^R \op P_{k+j}$ and $Z_z$ is the corresponding normalization constant. This state again admits a MPS representation \eqref{eq:MPS z} in which $\op{X}_0$ and $\op{X}_1$ are $(R+1)\times (R+1)$
matrices and $\vec{l}$ and $\vec{r}$ $(R+1)$-vectors \cite{Levi_2014}. As a final remark before describing the state preparation protocol, we would like to mention that the states (\ref{eq:ideal_z}) and (\ref{eq:ideal_zR}) are directly related to the ground states of the Hamiltonian (\ref{eq:full_hamil}), see Appendix \ref{app:MPS GS} for details.

In the following, we wish to show that a state of the form \eqref{eq:ideal_z} (i.e. $R=1$) can be generated via an appropriate sequence of pulses (see Appendix \ref{app:Generic R} for a more general procedure which applies to generic $R$). As a first step, we consider how a local pulse of area $A_k = \Omega_k t_k$ acts on a ground state atom located in site $k$. If the $k$-th atom is blockaded, i.e., if there is an excitation to its left and/or right, then
\begin{equation}
	\op U^{01}_k (A_k) \ket{0_k} = \ket{0_k},
\end{equation}
 whereas if it is not
 \begin{equation}
 	\op U^{01}_k \ket{0_k} (A_k) = \cos A_k \ket{0_k} + \sin A_k \ket{1_k}.
 	\label{eq:not_bl}
 \end{equation}
The excited component in $k$ will then blockade the following site, so that
 \begin{align}
 	\op U^{01}_{k+1} (A_{k+1}) \, \op U^{01}_k  (A_k) \ket{0_k 0_{k+1}}
 	& = \cos A_k \cos A_{k+1} \ket{0_k\,0_{k+1}} +\cos A_k \sin A_{k+1} \ket{0_k\, 1_{k+1}}   \nonumber \\
 	& +  \sin A_k \ket{1_k\,0_{k+1}}.
 \end{align}
Applying an ordered sequence of local pulses from the first to the last atom $\op U = \prod_{k=N}^1 \op U^{01}_k$ on the global atomic ground state $\ket{0_1 \ldots 0_N}$ will generically yield components on all elements of the basis in which no neighboring pairs of excitations appear. From Eq.~\eqref{eq:not_bl} it is not difficult to see that the component over the initial state will be $C_0 \equiv \prod_{k=N}^1 \cos A_k$. An excitation in, say, site $j$ will instead come with a $\sin A_j$ and a missing factor $\cos A_{j+1}$. Using this simple rule, we can work out that the ratio between the component of a generic basis element $C^{(\vec{n})}$ with $n$ (non-neighboring) excitations in $\vec{n}=\{j_1, \ldots j_n\}$ and $C_0$ will be given by
\begin{equation}
	\frac{C^{(\vec{n})}}{C_0} = \prod_{\mu = 1}^n \left(  \frac{\sin A_{j_\mu}}{\cos A_{j_\mu} \cos A_{j_{\mu} +1}}   \right).
\end{equation}
In order to correctly reproduce state $\ket{z}$, this ratio must be set to be equal to $z^n$. The only way for this to hold for every possible number $n$ of excitations is to have
\begin{equation}
	\frac{\sin A_{j}}{\cos A_{j} \cos A_{j +1}} = z \ \ \forall \, j.
	\label{eq:rec}
\end{equation}
This defines a recursion relation for $A_j$ in terms of $A_{j+1}$. The natural boundary condition to provide a seed to the recursion is
\begin{align}
  \cos(A_{N+j}) = 1\quad \forall \, j>0,
  \label{eq:MPS BC}
\end{align}
which corresponds to requiring that, should the excitation protocol stop before reaching the actual end of the chain, all atoms which have not been addressed should still be in their ground state. Equations \eqref{eq:rec} and \eqref{eq:MPS BC} can be analytically solved to yield
	\begin{align}
		\cos A_k = \sqrt{2\frac{\left( 1 + \sqrt{1+4z^2} \right)^{N+2-k} - \left( 1 - \sqrt{1+4z^2} \right)^{N+2-k}}{\left( 1 + \sqrt{1+4z^2} \right)^{N+3-k} - \left( 1 - \sqrt{1+4z^2} \right)^{N+3-k}}}.
		\label{eq:R1case}
	\end{align}
Together with the relation $\rm{sign} (\sin A_k) = \rm{sign} (z)$, the above expressions provide a unique way to extract the pulse areas $A_k$. Using the values so obtained in the protocol described in Table \ref{tab:GHZ_protocol} will yield state $\ket{z}$.


\change{\subsection{Quantum state transport}}
\label{sec:Teleportation ideal}

In this section we discuss a protocol for the coherent \change{transport} of a single qubit
state between the two ends of the chain, last column of Table \ref{tab:GHZ_protocol}. We consider a state $\ket{\psi_1}=\alpha \ket{0} + \beta \ket{1}$ to be initialized at the first qubit so that the total initial state reads
\begin{equation}
	\ket{\Psi_{\norml{in}}} = \left(\alpha \ket{0} + \beta \ket{1}\right)_1 \bigotimes \ket{0_2\, 0_3\, \ldots 0_N}.
\end{equation}
The circuit representation of the \change{state transport} protocol is shown in Fig. \ref{fig:MPS_n_teleport}b. The protocol relies on a sequence of three-body Toffoli gates (cf. Eq. (\ref{eq:hamil_lopt}) and Appendix \ref{app:Toffoli}), which, when applied at the two ends of the chain, becomes effectively a CNOT gate due to the absence of one of the sites. In our implementation, the Toffoli and the CNOT gate flip the target qubit if the controlled qubits are in the ground state $\ket{0}$.

The first step is the application of a $\pi$ pulse on the second qubit
\begin{align}
 \ket{\Psi_1} = \op{U}^{01}_{{2}}(\pi)\,\ket{\Psi_{\norml{in}}} = 
 \changem{\alpha \ket{0\, 1\, 0\, \ldots 0} + \beta \ket{1\, 0\, 0\, \ldots 0}}
\end{align}
As a second step, a $\pi$-pulse is applied on the first atom
\begin{align}
  \ket{\Psi_2} = \op{U}^{01}_{{1}}(\pi)\, \ket{\Psi_{\norml{1}}}
 =  \changem{ \alpha \ket{0\, 1\, 0\, \ldots 0} + \beta \ket{0\, 0\, 0\, \ldots 0}}\,,
\end{align}
which is then followed by the application of a $\pi$ pulse on the third qubit
\begin{align}
  \ket{\Psi_3} = \op{U}^{01}_{{3}}(\pi)\, \ket{\Psi_{\norml{2}}}
 = \changem{\alpha \ket{0\, 1\, 0\, \ldots 0} + \beta \ket{0\, 0\, 1\, \ldots 0}} \,.
\end{align}
and so on, cf. the steps (ii)-(iv) in Table \ref{tab:GHZ_protocol}, which are repeated till the end of the chain is reached.
At this stage, the initial state $\ket{\psi_1}$ has been successfully \change{transferred} so that the state of the $N$-th qubit reads
\begin{equation}
	\ket{\psi_N} = \begin{cases} 
    \norml{i}^{N+1} \op{\sigma}_y \ket{\psi_1} & N\;{\rm even} \\
       \ket{\psi_1} & N\;{\rm odd}.
   \end{cases}
\end{equation}
The presented state transport protocol requires $(2N-2)$ laser pulses and is deterministic as the only required information is the length of the spin chain with no need for a classical communication between the two ends of the chain \cite{a_Bennett_PRL_93}.\\

\section{Imperfections}
\label{sec:Imperfections}
The above presented protocols for {quantum state transport} and GHZ and dimer-MPS preparation rely on two important assumptions, namely that the blockade mechanism between two adjacent atoms excited to a Rydberg state is perfect and that the atoms are equally spaced in the chain. Here we analyze in detail the limitations of the protocols when these assumptions are relaxed, namely when the blockade becomes non-perfect due to the finite interaction strength \changer{between} nearest-neighbors {and when accounting for interactions beyond nearest-neighbors}, and due to the disorder in atomic positions coming from the finite temperature of the atoms and non-vanishing width of the optical traps. In the context of Rydberg gases, the disorder was shown to have a strong impact on the dynamics of spin excitations under the so-called ``anti-blockade'' conditions \cite{a_Marcuzzi_PRL_17}. We comment on other sources of imperfections, such as the influence of relaxation processes, at the end of this section. 
\subsection{Non-perfect Rydberg blockade}
\label{NON-Perfect_rydberg_blockade}

\change{The non-perfect blockade is accounted for by considering the full Hamiltonian (\ref{eq:full_hamil_piecewise}) featuring van der Waals interaction between all the atoms excited to a Rydberg state. In this case, when a pulse of area $A_k = \Omega_k\, t_k$ is applied to the $k-$th atom, the idealized unitary gate (\ref{eq:U 3body}) becomes
\begin{align}
 \op{W}^{01}_k(A_k = \Omega_k\, t_k) = \exp\left(\,-\norml{i}\,t_k \,\left( \op{h}_k\left(\Omega_k,\Delta_k = 0)\,+ \op H_\norml{int}\right)\right) \right) \, .
 \label{unit_full_hamil}
\end{align}
In what follows, considering the full Hamiltonian (\ref{eq:full_hamil_piecewise}) amounts simply to replacing the unitary gates $\op{U} \rightarrow \op{W}$ in Table \ref{tab:GHZ_protocol}, which are operators acting on the full Hilbert space of dimension $2^N$.} The non-perfect blockade results in configurations with adjacent Rydberg excitations. This limits the fidelity of the produced states defined as
\begin{align}
 F \equiv \bra{\Psi_\norml{target}} \rho_{\rm final} \ket{\Psi_\norml{target}}
 \label{Fid}
\end{align}
where $\rho_{\rm final}$ is the state at the output of the particular protocol described in Table \ref{tab:GHZ_protocol} and
$\ket{\Psi_\norml{target}}$ is the desired target state. We note that $\rho_{\rm final} = \ket{\Psi_\norml{final}}\bra{\Psi_\norml{final}}$ is pure by construction for the GHZ and dimer-MPS protocol as it is defined on all atoms of the chain; {on the other hand, it is generally mixed for the \change{transport} protocol as we trace over all but the last qubit}. In what follows we consider the Rabi frequency $\Omega$ to be identical for all atoms. {We also neglect the effect of the decay of the Rydberg states, which we justify in Sec. \ref{sec:Constraints}}.
\subsubsection{GHZ state preparation}
\label{GHZ-vdW}
\begin{figure}
\centering
  \includegraphics[width = 0.38\textwidth, angle= - 90]{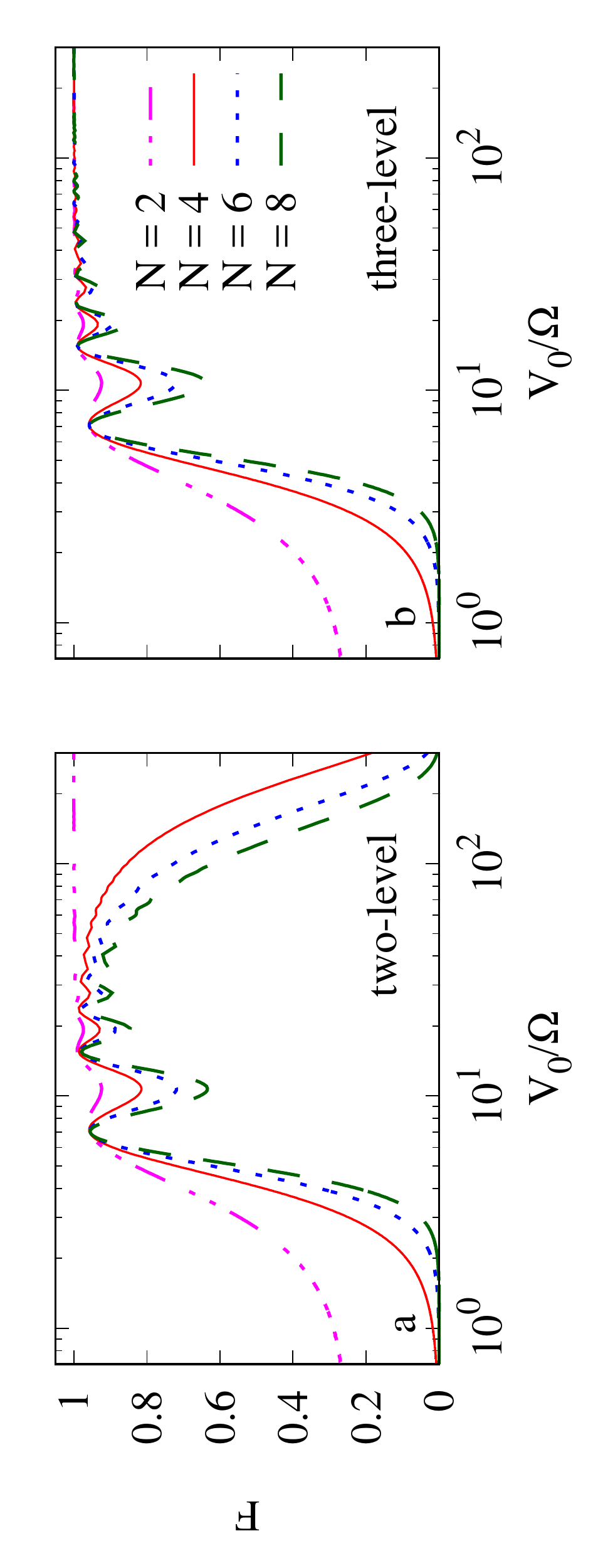}

   \caption{Fidelity of the GHZ state preparation protocol for a two-level description of the atoms
   (a), as well as the three-level scheme (b) as a function of $V_0/\Omega$. \changerE{In the three-level implementation of the protocol, atoms are excited to the Rydberg state to effectuate the blockade mechanism and are subsequently transferred to a stable hyperfine state, see text for details.} $F$ is given for chains of length $N~=~2,4,6,8$, respectively. 
  }
 
  \label{GHZ_fid}

\end{figure}

We start our analysis by examining the influence of the non-perfect Rydberg blockade on the GHZ state preparation described in the first column of Table \ref{tab:GHZ_protocol} with the replacement $\op{U}^{01} \rightarrow \op{W}^{01}$ and where $\ket{\Psi_\norml{target}} = \ket{\norml{GHZ}_{\rm N}}$ (we recall that here $N$ is even). The effect of the finite interaction strength can be seen in Fig. \ref{GHZ_fid}, where we plot the fidelity (\ref{Fid}) vs. the ratio of the nearest-neighbor interactions strength and the Rabi frequency $V_0/\Omega$. Fig. \ref{GHZ_fid}a shows a situation when using the two-level scheme. Fig. \ref{GHZ_fid}b then shows the fidelity in the three-level scheme, where the atoms are transferred from the Rydberg state $\ket{1}$ to the hyperfine state $\ket{\tilde{1}}$.

It is apparent from Fig. \ref{GHZ_fid} that in the limit $V_0/\Omega \rightarrow 0$, i.e. in the absence of the blockade, the fidelity goes to zero in all cases (except $N=2$ in the three-level scenario, where it is straightforward to show that $F \rightarrow 1/4$). In the opposite limit of the infinite blockade $V_0/\Omega \rightarrow \infty$, the vanishing of the fidelity in the two-level scenario can be easily understood as the blockade length extends over the whole chain allowing thus only for a single Rydberg excitation to be present. In contrast, in the three-level scenario there is never more than one atom in the Rydberg level at any given time, which results in unit fidelity in that limit.

For intermediate values of $V_0/\Omega$, one can observe oscillations of the fidelity $F$. The origin of those oscillations can be understood on the example of two atoms (cf. the magenta line in Fig. \ref{GHZ_fid}). Here the final state is obtained after application of one pulse at each atom and reads
\begin{align}
\ket{\Psi_{\rm final}} =\op W^{01}_2(\pi) \,  \op W^{01}_1 \left(\frac{\pi}{2}\right) \,  \ket{0 0} = \frac{1}{\sqrt 2} \left(  \ket{0\, 1} + \gamma \ket{1\, 0} + \delta \ket{\norml{1\,1}}\right)\,,
\label{GHZ_fid_2atoms}
\end{align}
where
\begin{subequations}
\begin{align}
 {\gamma} &= \norml{e}^{-\norml{i}\frac{\pi V_0}{8 \Omega}} \left(\cos\left( \pi\frac{\tau}{ {8 \Omega}} \,\right)
 +\frac{\norml{i} V_0\sin\left( \frac{\pi {\tau}}{8 \Omega}\,\right) }{{\tau}}\right) \\
 |\delta| &= \sqrt{1-|\gamma|^2} \\
 \tau &= \sqrt{ V_0^2 +16 \Omega^2}
 \end{align}
  \label{alpha_analytic}
\end{subequations}
The term $\delta\,\ket{1\,1}$ in (\ref{GHZ_fid_2atoms}) occurs due to the finiteness of the interaction strength $V_0$,
and reduces the fidelity of the produced state (\ref{Fid}), which reads
\begin{align}
 F_{N=2} = \frac{1}{4}\left| 1 + \gamma\right|^2\,.
 \label{eq:F GHZ}
\end{align}
While there are quantitative differences in the fidelity curves depending on the number of the atoms, we have verified that, for the atom numbers considered, the position of the maxima of the oscillations change only very slightly and thus the expressions (\ref{alpha_analytic}) provide accurate estimates (which become exact for $N=2$) for the values of $V_0/\Omega$ which maximize the fidelity.


 \subsubsection{Dimer-MPS preparation} 
 In analogy to the GHZ case, we investigate the effect of the non-perfect Rydberg blockade on the fidelity of the state obtained by applying the protocol described in the second column of Table \ref{tab:GHZ_protocol}, where $\ket{\Psi_{\rm final}} = \ket{z}$, Eq. (\ref{eq:MPS z}), and using again the replacement $\op{U}^{01} \rightarrow \op{W}^{01}$.
 \begin{figure}
 \centering
  \includegraphics[height=16cm, angle = -90]{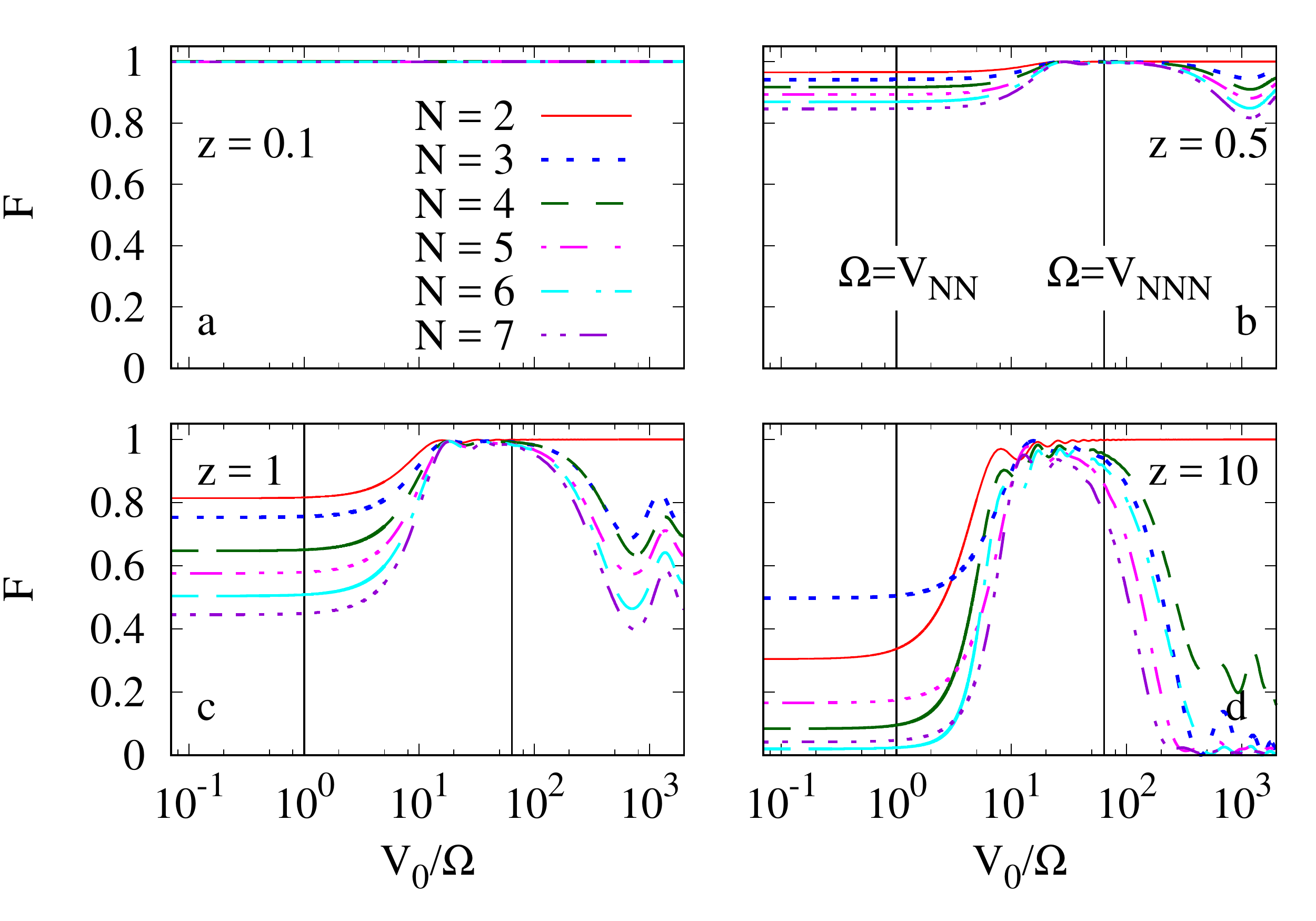}
  \caption{Fidelity of the dimer-MPS preparation protocol for $N~=~2$ up to $N~=~7$ atoms 
  as a function of $V_0/\Omega$ and for z=0.1,0.5,1,10 (a-d). See text for details.
  }
  \label{MPS_fid}
\end{figure}
Fig. \ref{MPS_fid} shows the influence of the non-perfect Rydberg blockade on the dimer-MPS state preparation protocol
on the example of $z~=~0.1,\,0.5,\,1,\,10$, for $N~=~2$ to $N~=~7$ atoms in a chain, respectively.
For $z~=~0.1$ (Fig. \ref{MPS_fid}a), the fidelity is approximately unity over the whole parameter range. This is a consequence of the fact that for $z \ll 1$, the dominant term in the state (\ref{eq:MPS z}) is the vacuum (all atoms in the ground state $\ket{0}$) which is not affected by the blockade mechanism.

When increasing $z$, terms containing more and more atoms in the Rydberg state $\ket{1}$ are becoming relevant with the limiting situation $z \gg 1$, where the dimer-MPS is dominated by terms containing $N/2$ [$(N+1)/2$] excitations for $N$ even [odd]. Similarly to the GHZ case, the drop of fidelity for small (large) $V_0/\Omega$ is due to non-perfect blockade (blockade extending beyond nearest-neighbor). 

We indicate by vertical lines in Fig. \ref{MPS_fid}b-d the values of $\Omega~=~V_{\rm NN}=V_0$ and $\Omega~=~V_{\rm NNN}$ corresponding to the interaction energies of nearest and next-to-nearest excitations respectively. The optimal fidelity can be expected to be found within the region delimited by these two boundaries. In fact, for $\Omega \gg V_{\rm NN}$ the blockade is relaxed and pairs of neighboring excitations are frequently produced. Conversely, for $\Omega \ll V_{\rm NNN}$ the next-to-nearest neighbor interaction (partially) blockades atoms two lattice sites away from an excitation, again spoiling the state preparation procedures described above.


\change{\subsubsection{Quantum state transport}}
\label{tele_vdW}

\begin{figure}
\centering
  \includegraphics[width=0.38\textwidth,angle = -90]{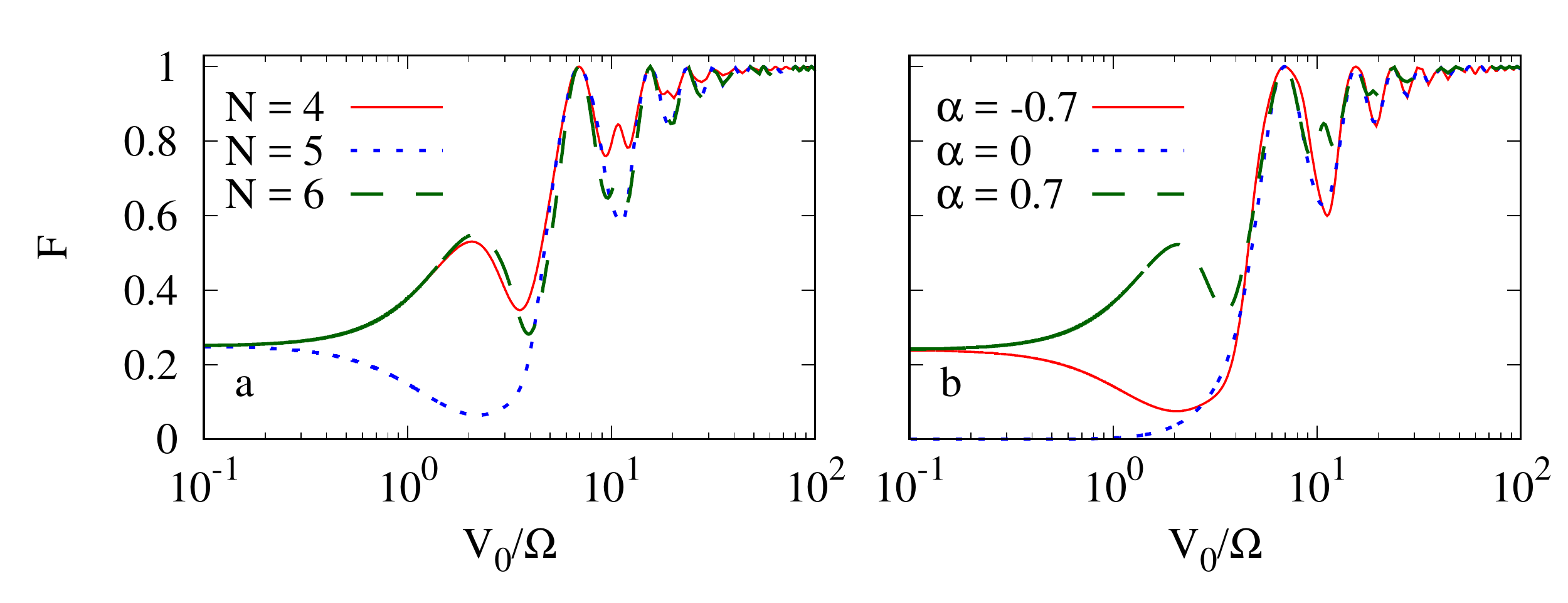}
  \caption{(a): Fidelity of a coherent \change{transport} process for $N~=~4,5,6$ as a function of $V_0/\Omega$. The \change{transferred} state is
  $\ket{\psi_1}~=~1/\sqrt{2}~(\ket{0}~+~\ket{1})$. (b): Comparison of the fidelity of the \change{transport} process for different initial states $\ket{\psi_1}$ for $N = 4$. The parameter $\alpha$ ($\beta~=~\sqrt{1-\alpha^2}$) of the state $\ket{\psi_1} = \alpha \ket{0} + \beta \ket{1}$ is varied.
   } 
     \label{fidelity_tele}
\end{figure}

Using the notation from Sec. \ref{sec:Teleportation ideal}, the target state is the state of the first atom to be \change{transferred}, $\ket{\Psi_{\rm target}} = \ket{\psi_1}$, while the final (single-qubit) state $\rho_{\rm final}={\rm Tr}_{\neq N} \left( \ket{\psi_t} \bra{\psi_t} \right)$ is obtained by tracing out all but the $N$-th atom and $\ket{\psi_t}$ is the target state obtained using the \change{transport} protocol with $\hat{U}^{01} \rightarrow \hat{W}^{01}$.
Fig. \ref{fidelity_tele}a shows the fidelity of the \change{transport} process for an initial state $\ket{\psi_1}~=~1/\sqrt{2}~(\ket{0}~+~\ket{1})$ at the first atom, for $N~=~4,5,6$. The decrease in the fidelity and the oscillations in the small and large $V_0/\Omega$ limit have the same origin as in the case of dimer-MPS and GHZ state preparation, namely a vanishing blockade for $V_0/\Omega \rightarrow 0$ and imperfect blockade due to the finiteness of $V_0$ respectively.

Next, in order to demonstrate \changerD{the influence of the initial state on the resulting fidelity}, in Fig. \ref{fidelity_tele}b we plot the fidelity vs. $V_0/\Omega$ for $\ket{\psi_1}~=~\alpha~\ket{0}~+~\beta~\ket{1}$, $\beta = \sqrt{1-|\alpha|^2}$, for $\alpha~=~-0.7,\,0,\,0.7$. It can be seen that in the parameter regime of interest ($V_0/\Omega \gtrsim 5$), the fidelity is independent of $\alpha$ with the maxima coinciding for the $\alpha$ considered. As in the GHZ case, the nature of the observed oscillations can be exemplified on the elementary example of two atoms for which we get
\begin{align*}
 \op W^{01}_1 (\pi)  \op W^{01}_2\left(\pi\right)\ket{\Psi_1}  &= \alpha \left(\, \gamma \ket{0\,1} + \gamma' \ket{1\,1}\right)
 +  \beta\left( \delta  \ket{0\,0} +\delta' \ket{1\,1} +\delta''\ket{0\,1}\right),
\end{align*}
where $\gamma, \delta$ are given by (\ref{alpha_analytic}) and
\changem{
\begin{subequations}
	\begin{align}
	        |\gamma'| &= \sqrt{1-|\gamma|^2}\\
%
%
%
		\delta' & = \frac{
	        2 \norml{e}^
		{- \frac{\norml{i} \pi V_0  }{4\Omega}} 
		\Omega
		\left(
		\ -\norml{i} V_0 
		+
		\norml{i} V_0
		\cos
		\left(
		   \frac{\pi \tau}{4 \Omega}
		\right)
		+ 
		\tau \sin 
		\left(
		\frac{\pi \tau}{4 \Omega}
		\right)
		\right)
		}
		{\tau^2}\\
		\delta'' &= \sqrt{1-|\delta|^2-|\delta'|^2}
	\end{align}
	\label{eq:Parameters tele}
\end{subequations}}
In the case of a perfect Rydberg blockade, only the terms $\alpha \gamma \ket{0\,1}$ and $\beta \delta \ket{0\,0}$ would occur after the application of the two laser pulses.

%
%
%
\subsection{Non-perfect blockade and position disorder}
\label{Disorder}

In lattice systems the atomic positions are typically considered to be fixed. In realistic experiments however, there is a finite uncertainty in their positions due to their finite temperature and finite extent of the optical traps realizing the lattice sites. This can have dramatic consequences e.g. on the excitation transport in the chains of Rydberg atoms \cite{a_Marcuzzi_PRL_17}. Based on our previous work \cite{a_Marcuzzi_PRL_17} we here recall the notions needed for the study of the effect of disorder on the protocol's performance.

The position disorder stems from the fluctuations of the atomic positions $\fe r_k = \fe r_k^{(0)} + \delta \fe{r}_k $ and subsequently the interaction energies (\ref{eq:Hint}). Here, $\delta \fe{r}_k $ are drawn from a three-dimensional Gaussian distribution
\begin{align}
 p_{\norml{pos}}(\fe r^{(k)}) = \frac{1}{(2\pi)^{2/3}\sigma_1\sigma_2\sigma_3}
 {\rm exp} \left[ -\frac{(\delta r_1^{(k)})^2}{2\sigma_1^2}-\frac{(\delta r_2^{(k)})^2}{2\sigma_2^2}-\frac{(\delta r_3^{(k)}-k r_0)^2}{2\sigma_3^2} \right] \,
 \label{Gaussian_distribution}
\end{align}
with widths $\sigma_i$, $i=1,2,3$ in the three spatial directions, see Fig. \ref{fig:scheme}a, where $\sigma_i$~=~$\sqrt{k_B T/(m \om_i^2)}$ is given by the temperature $T$ and the mass $m$ of the atoms and the trap frequencies $\om_i$ (see \cite{a_Marcuzzi_PRL_17} for details).

Taking for our comparison parameters from recent experiments \cite{a_Marcuzzi_PRL_17}, in the following we set the trap separation to $r_0 = 4.1 \;\mu {\rm m}$ and consider two scenarios for the disorder, one with isotropic disorder ($\sigma_i=120$ nm, $i=1,2,3$) and one with anisotropic disorder ($\sigma_1=1 \;\mu{\rm m}$ and $\sigma_{2,3}=120$ nm). The results presented below are obtained by averaging over 1000 realizations of the disorder unless stated otherwise.

\begin{figure}
\centering
  \includegraphics[width=0.38\textwidth,angle = -90]{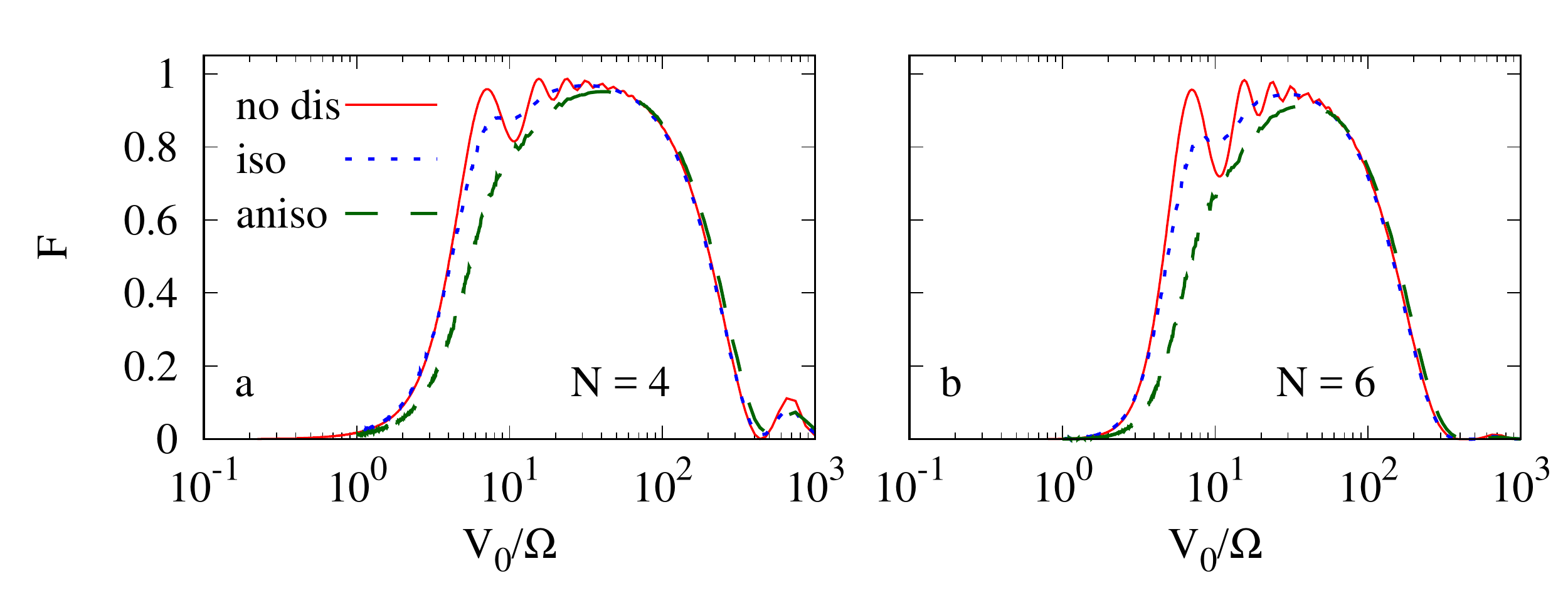}
  \caption{Fidelity of the GHZ state preparation protocol including the effect of non-perfect blockade and disorder for N = 4 (a) and N = 6 (b) atoms as a function of $V_0/\Omega$ with no (solid red), isotropic (dotted blue) and anisotropic (dashed green) disorder.
 }
\label{GHZ_fidelity_dis}
\end{figure}

\begin{figure}
\centering
  \includegraphics[height=16cm, angle= -90]{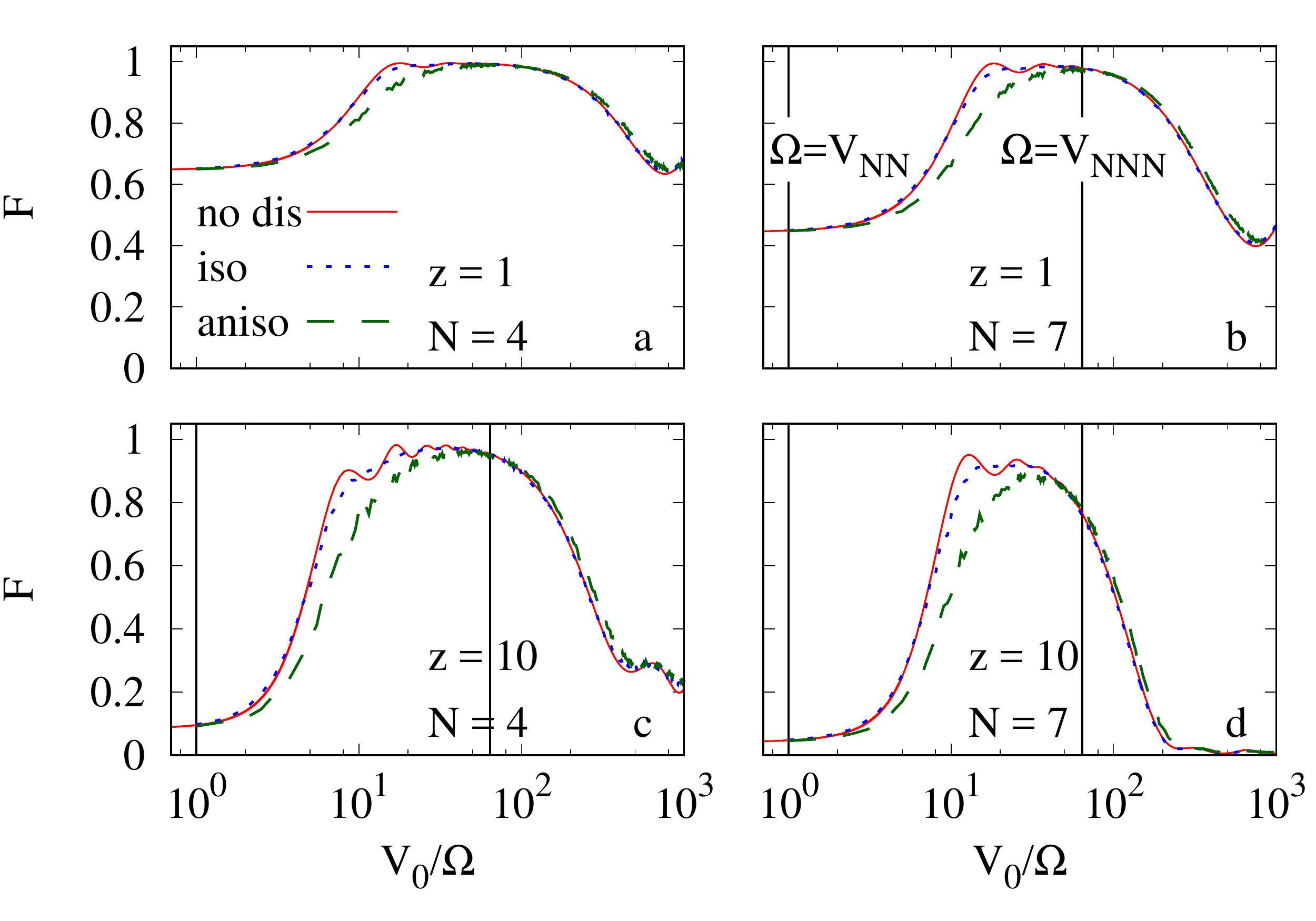}
  \caption{Fidelity of the dimer-MPS preparation protocol as a function of $V_0/\Omega$ including the effect of a non-perfect Rydberg blockade and disorder 
  for N = 4 (a,c), N = 7 (b,d) atoms and $z=1$ (a,b) and $z=10$ (c,d). The solid red, dotted blue and dashed green lines correspond to no, isotropic and anisotropic disorder respectively. Here, $F$ is obtained for 100 realisations of the disorder.}
\label{MPS_iso_dis}
\end{figure}

\begin{figure}
 \centering
  \includegraphics[height=16cm,angle = -90]{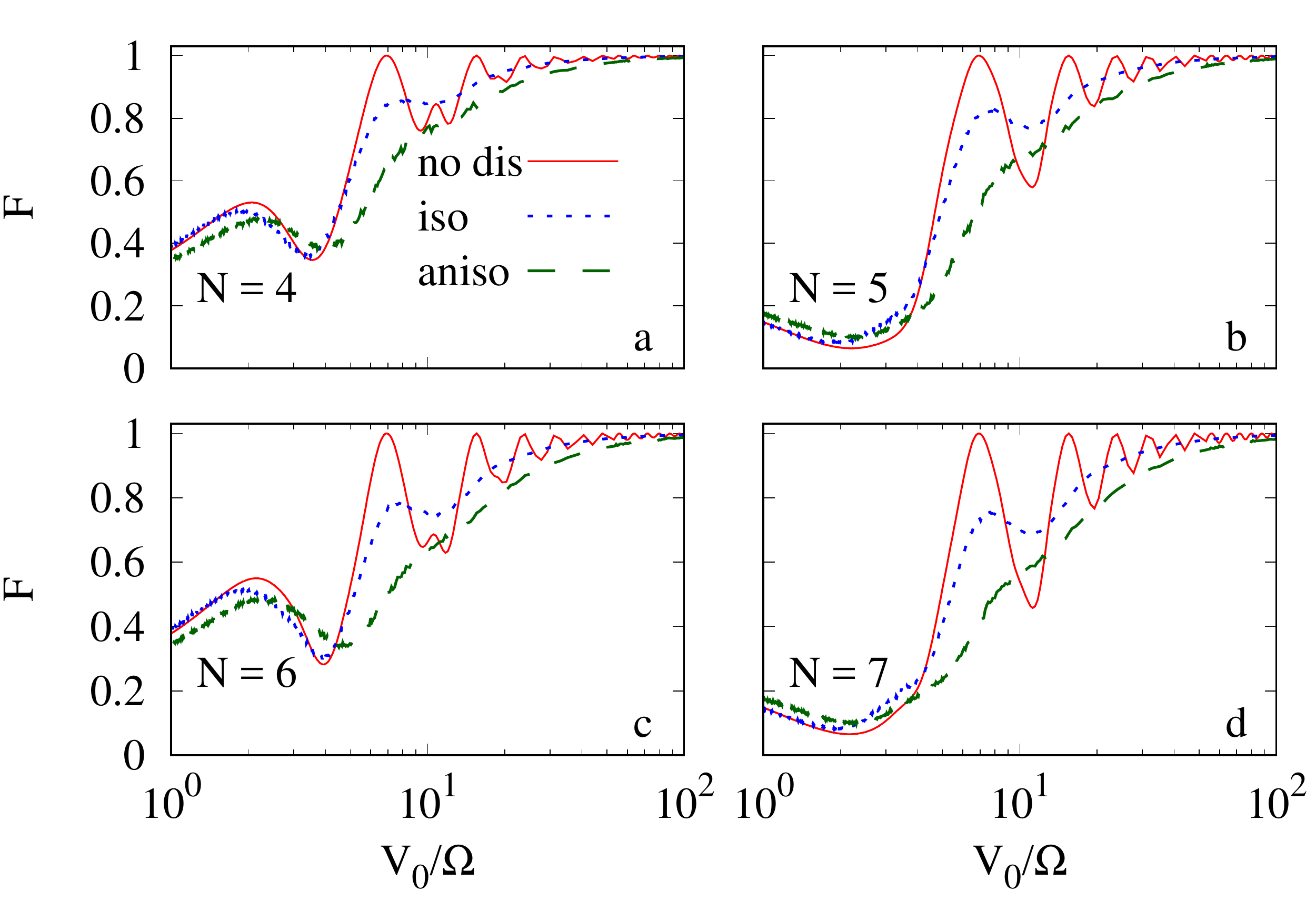}

  \caption{Fidelity of the \change{state transport} protocol as a function of $V_0/\Omega$ for the initial state $\ket{\psi_1}~=~1/\sqrt{2}~(\ket{0}~+~\ket{1})$ and for $N=4$ (a), $N= 5$ (b), $N=6$ (c), and $N=7$ (d). The solid red, dotted blue and dashed green lines correspond to no, isotropic and anisotropic disorder respectively.  
   }
\label{tele_iso}
\end{figure}

In Figs. \ref{GHZ_fidelity_dis}-\ref{tele_iso} we show the fidelity for the GHZ, dimer-MPS and \change{state transport} protocols respectively (we take the initial state $\ket{\psi_1} = 1/\sqrt{2} ( \ket{0} + \ket{1})$ in the \change{transport} protocol). 
{In all plots we compare lines corresponding to a non-perfect blockade and three different choices of disorder: absent (red), isotropic (dotted blue) and anisotropic (dashed green). }

The common feature to all plots is that, for the parameters considered, in the large $V_0/\Omega$ limit, the blockade mechanism dominates and is only weakly affected by the disorder: \changerE{here, all three lines show only small differences and sit well on top of each other}. Conversely, in the limit of $V_0/\Omega \sim O(1)$, which is of interest for \changerE{fast application of the protocols such that it still yields high-fidelity outputs (the larger the $\Omega$, the shorter the time needed to apply a pulse of a given area)}, the disorder has much stronger impact and in general decreases the fidelity significantly. In that regime, the decrease of fidelity is more pronounced with increasing disorder (situation in all Figs. \ref{GHZ_fidelity_dis}-\ref{tele_iso}) and also with increasing atom number (compare Fig. \ref{GHZ_fidelity_dis}a and Fig. \ref{GHZ_fidelity_dis}b, Fig.~\ref{MPS_iso_dis}a,c and Fig.~\ref{MPS_iso_dis}b,d and all the panels in Fig.~\ref{tele_iso}). The decrease of fidelity with the atom number stems from the fact that all our procedures address the atoms sequentially, and therefore each pulse, under imperfect conditions, will make the state diverge more from the target state. In the case of the dimer-MPS protocol, we note that the fidelity decreases also with increasing $z$ which can be easily understood as higher $z$ correspond to larger number of excitations in the state, which in turn is more sensitive to the disorder. 

The main message to be extracted from these results is that all the considered protocols are becoming more sensitive to the positional disorder when approaching the {fast operation regime $V_0/\Omega \sim O(1)$} from the large $V_0/\Omega$ side, where they are essentially insensitive to the amount of the disorder considered.

\begin{figure}
\centering

  \includegraphics[width=0.38\textwidth,angle = -90]{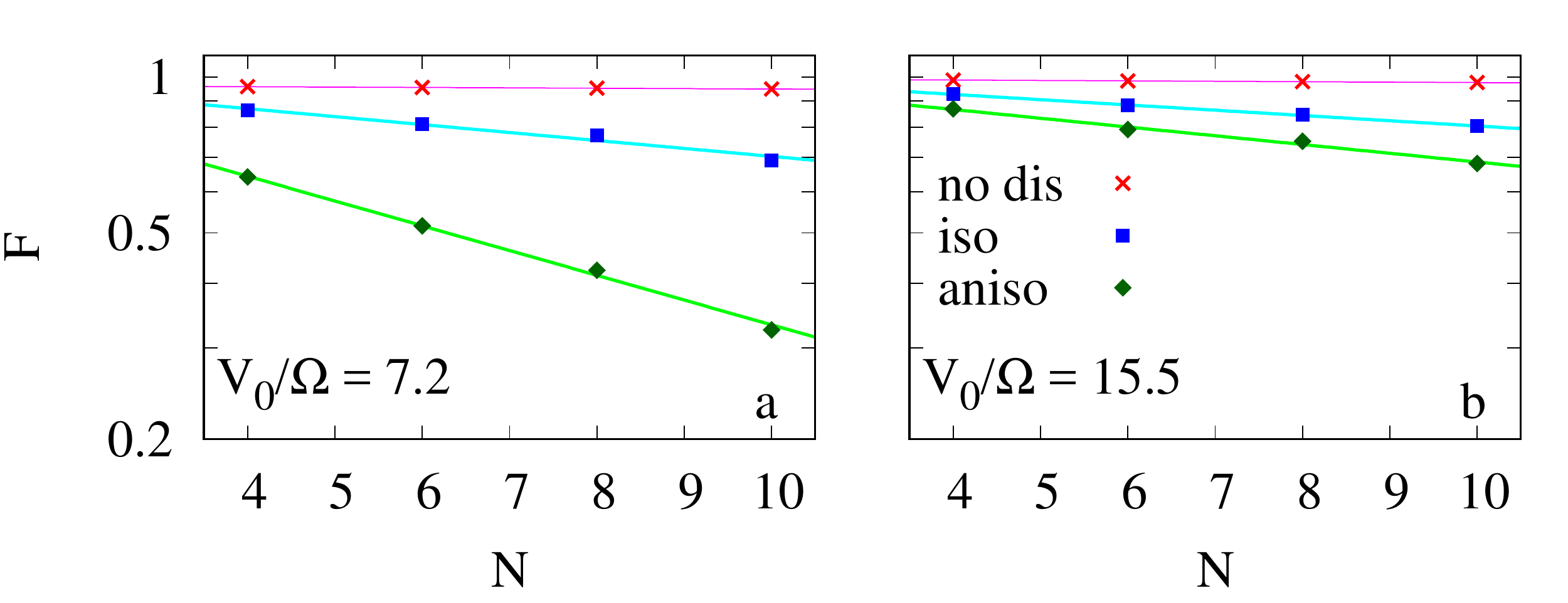}
  \caption{Fidelity of the \change{GHZ} protocol as a function of the number of atoms in the chain $N$
  for $V_0/\Omega$ = 7.2 (a) and  $V_0/\Omega$ = 15.5 (b). 
  The red crosses, blue squares and green diamonds  correspond
  to no, isotropic and anisotropic disorder respectively. The solid lines represent an exponential function fitted to the data points, see text for details. {\color{black} Data obtained by averaging over 100 realizations of the disorder.}
 }
\label{Fit_GHZ}
\end{figure}

\begin{figure}
\centering

  \includegraphics[width=0.5\textwidth,angle = -90]{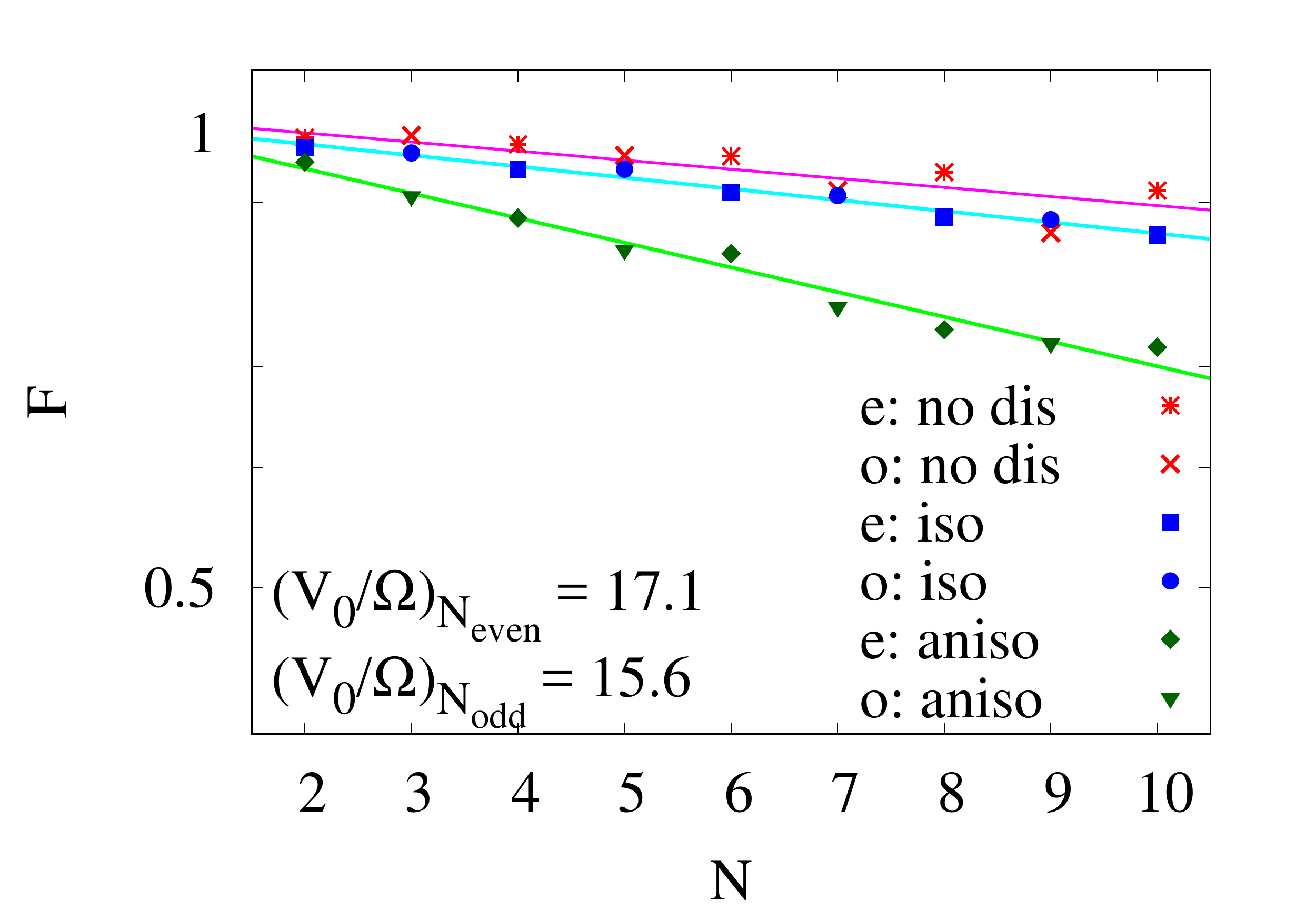}
   \caption{Fidelity of the \change{dimer-MPS} protocol as a function of the number of atoms in the chain $N$
  for $V_0/\Omega$ = 17.1 for even (e) and  $V_0/\Omega$ = 15.6 for odd (o) number of atoms in the chain and $z=10$. 
  The red crosses, blue squares and green diamonds  correspond
  to no, isotropic and anisotropic disorder respectively. The solid lines represent an exponential function fitted to the data points, see text for details. {\color{black} Data obtained by averaging over 100 realizations of the disorder.}
 }
\label{Fit_MPS} 
\end{figure}

\begin{figure}
\centering

  \includegraphics[width=0.38\textwidth,angle = -90]{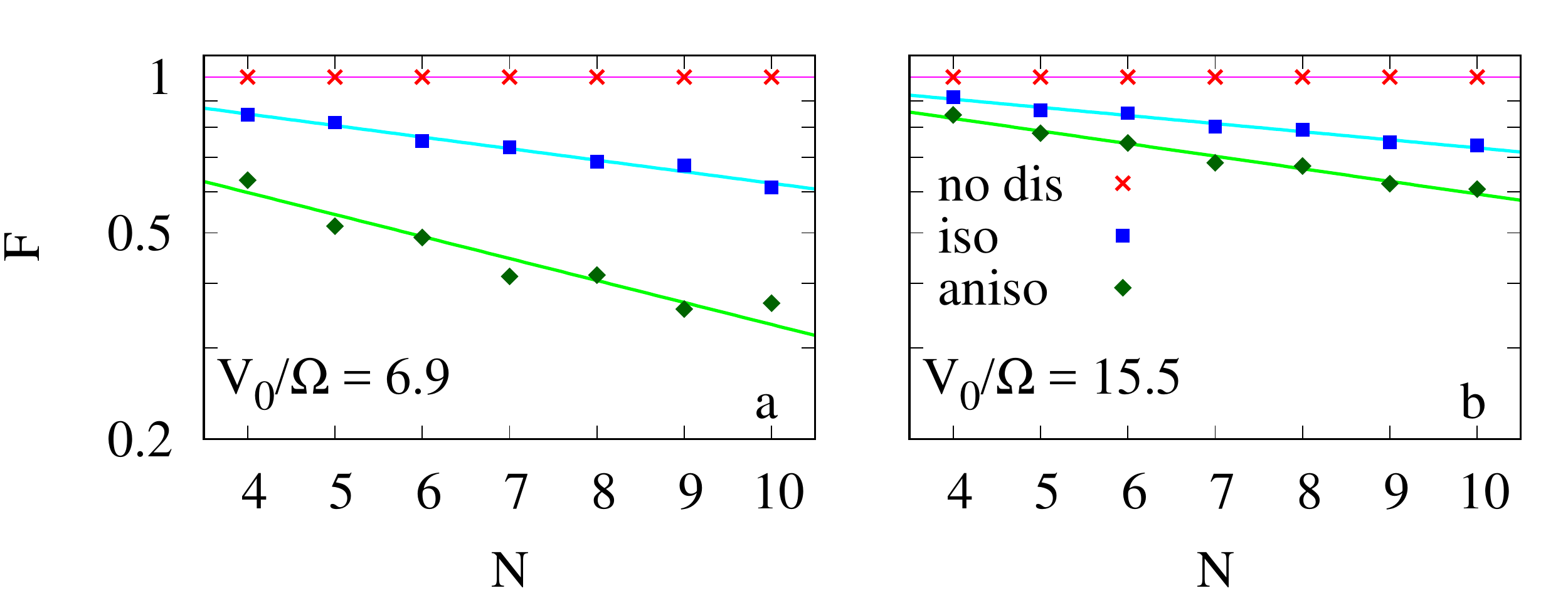}
  \caption{Fidelity of the \change{state transport} protocol as a function of the number of atoms in the chain $N$
  for $V_0/\Omega$ = 6.9 (a) and  $V_0/\Omega$ = 15.5 (b). 
  The red crosses, blue squares and green diamonds  correspond
  to no, isotropic and anisotropic disorder respectively. The solid lines represent an exponential function fitted to the data points, see text for details. {\color{black} Data obtained by averaging over 1000 realizations of the disorder.}
 }
\label{Fit}
\end{figure}

{\color{black}
In order to further quantify the sensitivity to the disorder and the corresponding decrease in fidelity, we study the fidelity as a function of the length of the chain for the GHZ, dimer-MPS state preparation and state transport protocols, which is shown in Figs. \ref{Fit_GHZ}, \ref{Fit_MPS} and \ref{Fit} respectively. Here we consider a system with non-perfect Rydberg blockade without disorder (red crosses), with isotropic (blue squares), and anisotropic (green diamonds) disorder. As we are interested in the fast application of the protocols, we have fixed the ratio $V_0/\Omega$ to 7.2 and 15.5 in Figs. \ref{Fit_GHZ}a,b and to $V_0/\Omega$ to 6.9 and 15.5 in Figs.\ref{Fit}a,b corresponding to the leftmost (second leftmost) peak in the fidelity, cf. Figs. \ref{GHZ_fid} and \ref{fidelity_tele} (the optimal values of $V_0/\Omega$ used can be extracted numerically or using (\ref{alpha_analytic}) and (\ref{eq:Parameters tele}) respectively). The solid lines correspond to an exponential function $f(N) = a \exp(-b(N-2))$ fitted to the respective data points.
Since we consider chains of length $N~\ge~2$ we set the exponent of the function $f(N)$ to $-b(N-2)$ rather than $-bN$. With such choice of the fitting function, the parameter $a$ in Table \ref{Fit_coeff} states the maximal fidelity of the protocol achievable in the simpliest possible system $N=2$ for different types of disorder. Table \ref{Fit_coeff} shows that the protocols do not reach a fidelity of one when disorder is considered. Figs. \ref{Fit_GHZ} and \ref{Fit} further quantify the above discussed observation that the resulting performance is a tradeoff between how fast the protocol can be applied and the resulting fidelity. Interestingly, in the absence of disorder and for $V_0/\Omega$ corresponding to the optimal fidelity regions, the final fidelity of the protocols is essentially insensitive to the exact atom number, {i.e. is not affected by the tails of the interaction potential}, see also Figs. \ref{GHZ_fid},\ref{fidelity_tele}.

Concerning the dimer-MPS state preparation protocol, in order to emphasize the effect of the long-range nature of the interactions on the resulting fidelity, we consider a dimer-MPS state with $z=10$, corresponding to the largest value we 
have considered here for this parameter. Our choice of $V_0/\Omega$ -- corresponding to the two leftmost peaks of the fidelity -- for the GHZ preparation and state transport protocols was motivated by the fact that in the absence of disorder these values of $V_0/\Omega$ provide a satisfactory tradeoff between the achievable fidelity and the speed of operation independently of the atom number. On the other hand, for the dimer-MPS protocol it is clear from Fig. \ref{MPS_fid} ($z=10$) that $V_0/\Omega$ corresponding to the leftmost peak of $N=2$ yields a rather rapid drop in the fidelity for higher $N$ even in the absence of the disorder. For this reason we consider only the values of $V_0/\Omega$ corresponding to the second leftmost peak of the fidelity in Fig. \ref{MPS_fid}. Here we note that the corresponding value of $V_0/\Omega$ slightly varies for odd and even $N$ unlike for the GHZ state preparation and transport protocols. Specifically, we find that $V_0/\Omega \approx 15.6$ and $V_0/\Omega \approx 17.1$ for odd and even $N$ which we use in Fig. \ref{Fit_MPS}.
}

\begin{table}[H]
  \begin{center}
    \begin{tabular}{|c|| c|c|c| c |c|c|}
        \hline
    &\multicolumn{4}{c|}{GHZ state}\tspace\\
    \hline
     &	\multicolumn{2}{c|}{$\frac{V_0}{\Omega}$ = 6.9}& \multicolumn{2}{c|}{$\frac{V_0}{\Omega}$ = 15.5}\tspace\\
     \hline
     \hline
     &   $a$                            & $b$              & $a$              & $b$\tspace\\
       \hline
no dis &   0.961015$\pm\,6\cdot10^{-6}$  & \quad (1739$\pm$ 1)$\cdot10^{-6}$ & 0.990331 $\pm\, 7\cdot10^{-6}$& \quad (1911 $\pm$ 1)$\cdot10^{-6}$ \tspace\\
iso    & ~ 0.93$\pm$ 0.02                & 0.035 $\pm$ 0.004                    &~ 0.970 $\pm$ 0.003               & 0.0234 $\pm$ 0.0006\tspace\\
aniso  &~ 0.80 $\pm$ 0.01                 &  0.110 $\pm$ 0.004                       & ~ 0.94 $\pm$ 0.01               & 0.039 $\pm$ 0.003\tspace\\ 
\hline
            \hline
    &\multicolumn{4}{c|}{dimer-MPS state}\tspace\\
    \hline
     &	\multicolumn{2}{c|}{$\left(\frac{V_0}{\Omega}\right)_{N\text{ odd}}$ = 15.6 and $\left(\frac{V_0}{\Omega}\right)_{N\text{ even}}$ = 17.1}& \tspace\\
     \hline
     \hline
     &   $a$                            & $b$              &              & \tspace\\
       \hline
no dis &   1.00$-\,0.02$  &            0.0139 $\pm$    0.004                   & & \tspace\\
iso    & ~ 0.983$\pm$ 0.004                & 0.0170 $\pm$ 0.0009                    &               &  \tspace\\
aniso  &~ 0.947 $\pm$ 0.009                 &  0.037 $\pm$ 0.002                       &          & \tspace\\ 
\hline
    \hline
    &\multicolumn{4}{c|}{State transport}\tspace\\
    \hline
     &	\multicolumn{2}{c|}{$\frac{V_0}{\Omega}$ = 6.9}& \multicolumn{2}{c|}{$\frac{V_0}{\Omega}$ = 15.5}\tspace\\
     \hline
     \hline
     &   $a$                            & $b$              & $a$              & $b$\tspace\\
       \hline
no dis &  0.999988 $\pm\, 4\cdot10^{-6}$  & \qquad ~(7.6 $\pm$ 0.8)$\cdot10^{-7}$ & 0.999989  $\pm\, 4\cdot10^{-6}$& \qquad ~~(6.8 $\pm$ 0.8)$\cdot10^{-7}$ \tspace\\
iso    & ~ 0.94 $\pm$ 0.05                & 0.051 $\pm$ 0.003                     &~ 0.97 $\pm$ 0.01               & 0.036 $\pm$ 0.002\tspace\\
aniso  &~ 0.73 $\pm$ 0.04                 & 0.09 $\pm$ 0.01                       & ~ 0.93 $\pm$ 0.02               & 0.055 $\pm$ 0.003\tspace\\ 
\hline
    \end{tabular}
  \end{center}
  \caption{\change{The coefficients of the function $f(N) = a\exp(-b\, (N-2))$ fitted to the data points in Figs.
  \ref{Fit_GHZ}, \ref{Fit_MPS} and \ref{Fit} for the GHZ, dimer-MPS state preparation and state transport protocols respectively. The abbreviations correspond to ``no dis'' = no disorder, ``iso'' = isotropic disorder, ``aniso'' = anisotropic disorder parameterized according to the values reported in the main text.}}
  \label{Fit_coeff}
\end{table}

\subsection{Limitations due to spontaneous decay}
\label{sec:Constraints}

In Sec. \ref{sec:Imperfections} A,B we have considered two sources of imperfections introducing errors in the state preparation and \change{state transport} protocols, namely the effect of the finiteness of the interaction potential resulting in non-perfect blockade and the effect of positional disorder of the atoms. We have neglected other noise sources such as spontaneous decay of the atoms from the Rydberg state or loss and dephasing mechanisms due to the interaction of the lattice atoms with the background gas. The reason for this is that we are interested in the short-time dynamics, where these effects become essentially negligible. To give a specific example of a constraint such considerations impose on the evolution of the system, we consider here the example of spontaneous decay.

Motivated by \cite{a_Marcuzzi_PRL_17}, we set the total duration of the experiment to $\tau_{\rm exp} = 2\;\change{\mu{\rm s}}$, which was used in \cite{a_Marcuzzi_PRL_17} in order to avoid effects due to the spontaneous photon emission, atom loss and mechanical effects induced by the forces between atoms. As we are considering non-adiabatic state-preparation protocols, we want to find a parameter regime where the protocol can be performed as fast as possible with a high fidelity.  Thus, we choose $(V_0/\Omega)^* = 6.9$, corresponding to the position of the leftmost maximum fidelity peak in each protocol (Figs. \ref{GHZ_fid}-\ref{fidelity_tele}). Finally, we fix $V_0=2\pi \times 8.4~$MHz \cite{a_Marcuzzi_PRL_17}. We can now estimate, with the help of the Table \ref{tab:GHZ_protocol} and the relations (\ref{eq:MPS BC}), the maximal number of atoms $N_{\rm max}$ so that the total duration of each protocol is smaller than $\tau_{\rm exp}$. The result can be summarized as
\begin{align}
 N_\norml{max} ~=~\left(\norml{GHZ}\rightarrow 9;~\norml{MPS}(z= 1) \rightarrow 13,\, {\rm MPS}(z= 10)\rightarrow 7;~ \norml{transport}\rightarrow 6\right).
\end{align}
One can see, that in this \change{specific $\tau_{\rm exp}=2\;\mu {\rm s}$} example, the protocols are limited to rather small number of atoms of the order of 10.
\change{In the case of the two-level scheme, where the final state contains Rydberg excitations, the coherence time of the system is dominated by  the lifetime $\tau_l$ of a single Rydberg atom
divided by the number of atoms $\tau_l/N$.} At the same time, Rydberg atoms provide multiple possibilities with relaxation timescales ranging over several orders of magnitude, typically from $\mu$s to ms regime \cite{Gallagher_1994,a_Saffman_RMP_10} depending on the Rydberg state. It would thus be interesting to identify the transitions with sufficiently large interaction strength $V_0$ and long relaxation times. That allows significantly higher $N_{\rm max}$, which can be in principle achieved by analyzing higher principal quantum number $n$ states (we recall that the interaction strength and relaxation timescales obey approximately the scaling relations $V_0 \propto n^{11}, \tau \propto n^3$ \cite{Low_2012}, while $n=56$ was used in \cite{a_Marcuzzi_PRL_17} in the repulsive interaction regime).
\change{However, when the three level scheme is used, where the excited Rydberg states are transferred to an atomic hyperfine ground state $\ket{\tilde 1}$,
the coherence time can be seen to last up to several tens of seconds as outlined in reference \cite{a:Szmuk_15}.}

\section{Conclusions}
\label{sec:conclusion}

{We have described three different protocols for quantum information processing based on non-adiabatic manipulation of atoms. The protocols exploit the Rydberg blockade mechanism and require single site addressability, which is now an available experimental tool. Specifically, we have shown how to generate antiferromagnetic GHZ states, a class of matrix product states - the dimer-MPS - which include the Rydberg crystals, and \change{quantum state transport} in chains of Rydberg atoms.} We have evaluated the effect of the full interaction potential on the quality of the protocols identifying a parameter regime yielding optimal performance in terms of fast operation resulting in output states with high fidelity. Next, we have evaluated the experimentally relevant effect of positional disorder stemming from the finite temperature of the atoms and width of the optical traps respectively, and quantified its influence on the fidelity. Finally, we have discussed the constraints imposed on the presented protocols by other sources of imperfections, namely the relaxation of the Rydberg states.

\section{Acknowledgement}

\changerE{We would like to thank Daniel Barredo, Antoine Browaeys, Thierry Lahaye, Sylvain de L\'{e}s\'{e}leuc and Markus M\"{u}ller for fruitful discussions.} The research leading to these results has received funding from the European Research Council under the European Union's Seventh Framework
Programme (FP/2007-2013) / ERC Grant Agreement No. 335266 (ESCQUMA), the EU-FET Grant No. 512862 (HAIRS), the H2020-FETPROACT-2014 Grant No.
640378 (RYSQ), and EPSRC Grant No. EP/M014266/1.

\bibliography{bibliography}
  \bibliographystyle{unsrt}        

\appendix

\section{Toffoli gate}
\label{app:Toffoli}

\changerE{The basic building block of the protocols studied in this work is the three-body Hamiltonian (\ref{eq:hamil_lopt}) and the associated unitary gate (\ref{eq:U 3body}).
The latter provides a tool to implement various three-body gates,
where the specific properties of the gate are determined by the parameters of the Hamiltonian,
namely the detuning $\Delta_k$, Rabi frequency $\Omega_k$ and the duration of the pulse $t_k$ on $k$-th atom.
In the specific case, where $\Delta_k = 0$ for all atoms and the area of the pulse applied at the $k$-th atom $\Omega_k t_k = \pi$, the unitary (\ref{eq:U 3body}) corresponds to the Toffoli gate.} 

Here, in Table \ref{tab:Truth_toffoli}, we list for completeness the properties of the Toffoli gate used in the \change{quantum state transport} protocol.
The first and last qubits are the control ones while the second qubit is the target.
Provided that both control qubits are in state $\ket{0}$ the Toffoli gate acts like a $\pi$-pulse on the target qubit.

\begin{table}[H]
  \begin{center}
    \begin{tabular}{c| c}
	Input& Output\tspace\\
	\hline
	000  & 010\tspace\\
	001  & 001\\
	010  & 000 \\
	011  & 011 \\
	100  & 100 \\
	101  & 101 \\
	110  & 110 \\
       111  & 111
    \end{tabular}
  \end{center}
  \caption{Truth table of the Toffoli gate. The first and the third qubits are the control qubits, while the second qubit is the 
  target. When the control qubits are both in the ground state $\ket{0}$, the Toffoli gate corresponds to an application of a $\pi$-pulse on the target qubit.}
  \label{tab:Truth_toffoli}
\end{table}
\section{On the derivation of the effective Hamiltonian}
\label{APP_LOPT}
  To the lowest order in the perturbation, the effective Hamiltonian is obtained by applying the unitary transformation
 $\op{U}= \exp\left({\norml{i}\,t\,V\,\sum_k^N\, \op{n}_k\, \op{n}_{k+1}}\right)$ on (\ref{eq:full_hamil}). Using the rotating wave approximation, this
leads to
\begin{align}
 \op{H}_{\norml{int}} = \Omega\, \sum_{k=1}^N\, \op{\sigma}_y^{k}\, (1-\op n_{k-1})(1-\op n_{k+1}) +\frac{V_0}{64} \,\sum_{k=1}^N \, \op n_k\, \op n_{k+2}
 \label{eq:H lopt}
\end{align}
Next, the second term in (\ref{eq:H lopt}) is suppressed by a factor 64 and is neglected, which yields the effective Hamiltonian (\ref{eq:hamil_lopt}).

\section{Dimer-MPS}
\label{app_RK}
\subsection{Derivation of the recurrence formulae}

We discuss here how to retrieve a recursion formula for the areas $A_k$ in the case where an excitation blocks its first $R$ neighbors to the right and to the left. Formula \eqref{eq:rec} will correspond to the particular case $R=1$. We start by recalling that the target state is expressed as
\begin{align}
	\ket{z}  = \frac{1}{\sqrt{Z_z}}  \prod_{k=1}^{L}
	\left( \mathbb{1} + \,z\, \op{P}_{k, {\rm left}} \,\op{\sigma}_k^+ \, \op{P}_{k, {\rm right}} \right)  \ket{\downarrow \ldots \downarrow},
	\label{eq:state}
\end{align}
where $\op{P}_{k, {\rm left}} = \op P_{k-1} \ldots \op P_{k-R}$ and $\op P_{k, {\rm right}} = \op P_{k+1} \ldots \op P_{k+R}$. The action of a local pulse on the $k$-th atom depends on whether the latter is blockaded (i.e.~there are excitations within a radius $R$) or not. In the former case, we have
\begin{equation}
	\op U^{01}_k (A_k) \ket{0_k} = \ket{0_k}.
	\label{eq:blx}
\end{equation}
Conversely, if the atom is not blockaded,
 \begin{equation}
 	\op U^{01}_k \ket{0_k} (A_k) = \cos A_k \ket{0_k} + \sin A_k \ket{1_k}.
 	\label{eq:not_blx}
 \end{equation}
We now think of applying an ordered sequence of these unitary operations addressing one atom at a time, from the first to the last $\op U = \prod_{k=N}^1 \op U^{01}_k$, to a ground-state configuration $\ket{0_1 \ldots 0_N}$. This will yield a state with components on all configurations in which no pairs of excitations appear at a distance $\leq R$. Note that, according to \eqref{eq:not_blx}, the ground state configuration will come with a coefficient
\begin{equation}
	C_0 \equiv \prod_{k=N}^1 \cos A_k.
\end{equation}
Every configuration with an excitation in site $j$ will instead feature a factor $\sin A_j$. Furthermore, because of the blockade, the operations acting on $j = k+1, \ldots ,k+R$ will behave as displayed in \eqref{eq:blx}, i.e.~they will trivially contribute $1$ to the overall coefficient. In general, we can reconstruct the coefficient of a generic configuration via the following simple rules:
\begin{itemize}
	\item[(i)] Choose a configuration and start reading it from the first atom to be addressed to the last one. Assign a coefficient $1$ to start with.
	\item[(ii)] Until a $\ket{1}$ is found, for every $\ket{0_k}$ multiply the coefficient by $\cos A_k$.
	\item[(iii)] When a $\ket{1_k}$ is found, multiply by $\sin A_k$ and skip to position $k+R + 1$.
	\item[(iv)] Apply (ii) again.
\end{itemize}
Hence, if we call $C^{(1)}_j$ the coefficient of the configuration with a single excitation in site $j$ we have
\begin{equation}
	\frac{C^{(1)}_j}{C_0} = \frac{\sin A_j}{\prod_{k=j}^{j+R} \cos A_k}.
\end{equation}
 More in general, the coefficient $C^{(\vec{n})}$ of an allowed state with $n$ excitations in positions $\vec{n}= \{j_1, \ldots j_n\}$ will obey
 \begin{equation} 
 	\frac{C^{(\vec{n})}}{C_0} = \prod_{\mu=1}^n \frac{\sin A_{j_\mu}}{\prod_{k={j_\mu}}^{{j_{\mu}}+R} \cos A_k}.
 \end{equation}
The correct form of state $\ket{z}$ is then reproduced by fixing the areas in such a way that
   \begin{equation}
	\frac{\sin A_j}{\prod_{k=j}^{j+R} \cos A_k} = z \ \ \forall \, z.
   \end{equation}
This defines a recursion for $A_j$ in terms of $A_{j+1}, \ldots, A_{j+R}$. In order to make the solution of the recursion unique, we need to also impose the boundary conditions
\begin{equation}
	\cos A_{N+j} = 1 \ \ \forall \,\,\, 1 \leq j \leq R.
\end{equation}
Note that the recursion relation can be also cast in the form
\begin{equation}
	\tan{A_j} = z \prod_{k=j+1}^{j+R} \cos A_k.
\end{equation}
Since the $\tan$ function is $\pi$-periodic, we have the freedom to choose the sign of the cosines, which we determine to be all positive. This means that the sign of the sines will instead coincide with the sign of the parameter $z$. This uniquely identifies the areas $A_k$ modulo $2\pi$.

\subsubsection{Nearest-neighbor case}
	
	Here, we solve the recursion relation (\ref{eq:rec}) in the simplest case $R = 1$. By defining
	\begin{align}
		x_k = \cos^2 A_{L+1-k}
	\end{align}
	and $a = z^2$ we find
	\begin{align}
		x_{k+1} = \frac{1}{1 + a\, x_k}
	\end{align}
	with seed $x_0 = 1$. We now rewrite $x_k = p_k / q_k$, which yields
	\begin{align}
		\frac{p_{k+1}}{q_{k+1}} = \frac{q_k}{q_k + a\, p_k}
	\end{align}
	and assume that we can separate numerator and denominator as if they were independent, leading to the system
	\begin{align}
			p_{k+1} &= q_k \\
			q_{k+1} &= q_k + a\, q_{k-1}.
	\end{align}
	In order to correctly reproduce the boundary condition for $x_k$, we ask $q_0 = q_{-1} = p_0 = 1$. Since $a \geq 0$, we see that if $q_k > 0$ and $q_{k-1} > 0$, then $q_{k+1} > 0$ as well. Given the initial conditions, it follows from induction that $q_k > 0$ $\forall k$. Furthermore,
\begin{equation}
	q_{k+1} \geq q_k = p_{k+1},
\end{equation}
as expected since by definition $x_{k} = p_k / q_k = q_{k-1} / q_k$ must be $\leq 1$.

	The recursion equation for $q_k$ is linear and can be solved exactly: the associated polynomial is
	$\lambda^2 - \lambda - a$, whose roots are
	\begin{align}
		\lambda_{\pm} = \frac{1 \pm \sqrt{1 + 4a}}{2}.
	\end{align} 
	Therefore, the general solution is
	\begin{align}
		q_{k} = A \lambda_+^k + B \lambda_-^k,
	\end{align}
	with $A$ and $B$ fixed via the boundary conditions $q_{-1} = q_0 = 1$, which yield
	\begin{align*}
			A &=  \frac{1 + 2a + \sqrt{1+4a}}{2\sqrt{1+4a}} =  \frac{1}{\sqrt{1+4a}} \left( \frac{1 + \sqrt{1+4a}}{2} \right)^2 \\
 			B &= \frac{-1 - 2a + \sqrt{1+4a}}{2\sqrt{1+4a}}  =  - \frac{1}{\sqrt{1+4a}} \left( \frac{1 - \sqrt{1+4a}}{2} \right)^2.
	\end{align*}
	Thereby, we find
	\begin{align}
		q_k = \frac{1}{\sqrt{1+4a}}   \left[  \left( \frac{1 + \sqrt{1+4a}}{2} \right)^{k+2}  - \left( \frac{1 - \sqrt{1+4a}}{2} \right)^{k+2}  \right]
	\end{align}
	and, consequently,
	\begin{align}
		x_k& = \frac{p_k}{q_k} =
		\frac{\left( \frac{1 + \sqrt{1+4a}}{2} \right)^{k+1}  - 
		\left( \frac{1 - \sqrt{1+4a}}{2} \right)^{k+1} }{\left(t \frac{1 + \sqrt{1+4a}}{2} \right)^{k+2}  - 
		\left( \frac{1 - \sqrt{1+4a}}{2} \right)^{k+2} } \notag\\
		 &= 2\frac{\left( 1 + \sqrt{1+4a} \right)^{k+1} - \left( 1 - \sqrt{1+4a} \right)^{k+1}}{\left( 1 + \sqrt{1+4a} \right)^{k+2} - \left( 1 - \sqrt{1+4a} \right)^{k+2}}.
	\end{align}
	which exactly corresponds to Eq.~\eqref{eq:R1case}, since we have chosen the cosines to be positive.
%
%
%
\subsubsection{Generic $R$ case}
\label{app:Generic R}

The general case is defined by
\begin{align}
	x_{k+1} = \frac{1}{1 + a \prod\limits_{j=k}^{k+1-R} x_j}.
\end{align}
The same trick above $x_k = p_k/q_k$ can be applied, yielding
\begin{align}
	\frac{p_{k+1}}{q_{k+1}} = \frac{\prod\limits_{j=k}^{k+1-R} q_j}{\prod\limits_{j=k}^{k+1-R} q_j + a \prod\limits_{j=k}^{k+1-R} p_j}.
\end{align}
The same choice $p_k = q_{k-1}$ gives now
\begin{align}
	q_{k+1} = q_k + a q_{k-R},
\end{align}
which is associated to the polynomial
\begin{align}
	\lambda^{R+1} = \lambda^R + a,
\end{align}
which is generally not analytically solvable. However, it has $R+1$ (complex) roots $\lambda_j$ and the general implicit solution of the recursion is
\begin{align}
	q_k = \sum_{j=1}^{R+1} A_j \lambda_j^k. 
\end{align}
Imposing the boundary conditions yields a system of equations for the coefficients $A_j$ with equations of the form
\begin{align}
	\sum_{j=1}^{R+1} \frac{A_j}{\lambda_j^n} = 1,
\end{align}
for $n=-R, ... ,0$ which can be also written as
\begin{equation}
	\sum_{j=1}^{R+1} M_{nj} A_j = 1 \ \ \forall \, n,
\end{equation}
where $M_{nj}$ is a matrix with entries $M_{nj} = \lambda_j^n$, i.e.~it is a Vandermonde matrix. By using Cramer's rule for the solution of linear systems of equations, we can write the coefficient $A_k$ as
\begin{equation}
	A_k = \frac{ \det M^{(k)}}{\det M},
\end{equation}
where $M^{(k)}$ is constructed from $M$ by substituting $1$ to all entries on the $k$-th column. Note that $M^{(k)} = M$ if $\lambda_k=1$, which means that $M^{(k)}$ is also a Vandermonde matrix. Exploiting the known structure of the determinants of Vandermonde matrices we can thus write
\begin{align}
	A_k =  \frac{ \prod\limits_{\substack{1 \leq i < j \leq R+1}} \left( \frac{1}{\lambda^{(k)}_j} - \frac{1}{\lambda^{(k)}_i}  \right)}{ \prod\limits_{1 \leq i < j \leq R+1} \left( \frac{1}{\lambda_j} - \frac{1}{\lambda_i}  \right) },
\end{align}
where $\lambda_{j}^{(k)} = \lambda_j$ if $j \neq k$ and $1$ otherwise. By simplifying all common factors, this can also be rewritten as
\begin{align}
	A_k =  \lambda_k^R\, \frac{\prod\limits_{j \neq k}(1- \lambda_j)}{\prod\limits_{j \neq k}(\lambda_k - \lambda_j)}.
\end{align}

\subsection{Comment on the dimer-MPS as a ground state of the Rydberg Hamiltonian}
\label{app:MPS GS}

In \cite{a_Lesanovsky_PRL_11} it has been demonstrated that, in a specific regime of the parameters, the Hamiltonian \eqref{eq:full_hamil} can be approximately mapped to a Rokhsar-Kivelson form \cite{a_Rokhsar_PRL_88}, which is associated with the stochastic matrix describing the evolution of a classical stochastic process \cite{a_Castelnovo_2005}. These Hamiltonians admit an exact solution for their ground states. In particular, as shown in \cite{a_Lesanovsky_PRL_11}, Hamiltonian \eqref{eq:full_hamil} admits such a representation when its parameters satisfy $\Omega_k~=~\Omega$, $\forall k$, and
\begin{equation}
	\Delta_k \equiv \Delta = 2^6\, \frac{\Omega^2}{V_0} - \frac{3V_0}{2^6}.
\end{equation}
which identifies a manifold in the parameter space. Under these conditions, and assuming that interactions among nearest neighbors are strong enough to enforce an almost perfect blockade, while becoming sufficiently small beyond next-nearest neighbors to be safely neglected, the ground state of the system is well approximated by (\ref{eq:ideal_z}), where $z = -V_0 / (2^6 \Omega)$.

Next, as mentioned in the main text, the same construction can be extended to the case of the blockade effect extending beyond nearest neighbors: assuming that an excitation prevents its first $R$ neighbors from being excited, the many-body ground state can be again analytically expressed in the parameter manifold \cite{Levi_2014}
\begin{equation}
	\Delta = \frac{\Omega^2 }{V_R} - V_R\, (2\,R+1),
\end{equation}
where $V_R = C_6 / (r_0^6 (R+1)^6)$, and takes the generalized form (\ref{eq:ideal_zR}).

\end{document}